\documentclass{aa}

\usepackage{color}
\usepackage{graphicx}
\usepackage{natbib}
\usepackage{epstopdf}
\usepackage[varg]{txfonts}
\begin{document}

   \title{Absolute parameters for AI Phoenicis using WASP photometry}%\thanks{}}

%   \subtitle{I. Overviewing the $\kappa$-mechanism}

   \author{J.~A.~Kirkby-Kent\inst{1}
          \and
          P.~F.~L.~Maxted\inst{1}
          \and
         {A.~M.~Serenelli}\inst{2}
          \and
          O.~D.~Turner\inst{1}
          \and
          D.~F.~Evans\inst{1}
          \and
          D.~R.~Anderson\inst{1}
          \and
          C.~Hellier\inst{1}
         \and
         R.~G.~West\inst{3}
          }

   \institute{Astrophysics Group,  Keele University, Keele, Staffordshire ST5
5BG, UK\\
              \email{j.a.kirkby-kent@keele.ac.uk}
 	\and 
	Institute of Space Sciences (ICE/CSIC-IEEC), Carrer de Can Magrans S/N, E-08193, Cerdanyola del Valles, Spain
	\and
	Department of Physics, University of Warwick, Coventry, CV4 7AL, UK}

   \date{Received ; accepted }

% \abstract{}{}{}{}{} 
% 5 {} token are mandatory
 
  \abstract
   {AI Phe is a double-lined, detached eclipsing binary, in which a K-type sub-giant star totally eclipses its main-sequence companion every 24.6 days. This configuration makes AI Phe ideal for testing stellar evolutionary models. Difficulties in obtaining a complete lightcurve mean the precision of existing radii measurements could be improved.}
   {Our aim is to improve the precision of the radius measurements for the stars in AI Phe using high-precision photometry from the Wide Angle Search for Planets (WASP), and use these improved radius measurements together with estimates of the masses, temperatures and composition of the stars to place constraints on the mixing length, helium abundance and age of the system.}
   {A best-fit \textsc{ebop} model is used to obtain lightcurve parameters, with their standard errors calculated using a prayer-bead algorithm. These were combined with previously published spectroscopic orbit results, to obtain masses and radii. A Bayesian method is used to estimate the age of the system for model grids with different mixing lengths and helium abundances.}
   {The radii are found to be \mbox{$R_{1}=1.835\pm0.014\,{\rm R}_{\sun}$}, \mbox{$R_{2}=2.912\pm0.014\,{\rm R}_{\sun}$} and the masses \mbox{$M_{1}=1.1973\pm0.0037\,{\rm M}_{\sun}$}, \mbox{$M_{2}=1.2473\pm0.0039\,{\rm M}_{\sun}$}. From the best-fit stellar models we infer a mixing length of 1.78, a helium abundance of \mbox{$Y_{\rm AI}=0.26^{+0.02}_{-0.01}$} and an age of \mbox{$4.39\pm0.32$\,Gyr}. Times of primary minimum show the period of AI Phe is not constant. Currently, there are insufficient data to determine the cause of this variation. }
  { Improved precision in the masses and radii have improved the age estimate, and allowed the mixing length and helium abundance to be constrained. The eccentricity is now the largest source of uncertainty in calculating the masses. Further work is needed to characterise the orbit of AI Phe. Obtaining more binaries with parameters measured to a similar level of precision would allow us to test for relationships between helium abundance and mixing length.
  
   }
   \maketitle
%
%________________________________________________________________

\section{Introduction}
\label{sec:Intro}

Stellar evolutionary models are used in a variety of areas of astrophysics, from predicting the properties of stars in galaxy formation models \citep{2013ASSL..396..345S}, to characterising planetary host stars \citep{2015MNRAS.447..846B, 2015A&A...579A..36C}. However, there are an increasing number of cases where the models have failed to reproduce the observed parameters of low-mass stars within binary systems. For example, observations of \object{IM Vir} (a G7+K7-type binary) found the radii of the primary and secondary components were larger than those predicted by the models by 3.7\% and 7.5\%, respectively \citep{2009ApJ...707..671M}, while the temperatures of the primary and secondary were found to be 100\,K and 150\,K (respectively) cooler than model predictions. \citet{2012A&A...540A..64V} showed the secondary of \object{EF Aqr} (a G0-type system) is 9\% larger and 400K cooler than model predictions. A similar situation was found for \object{V530 Ori} (a G1+M1-type binary), for which models predicted a radius 3.7\% smaller than observations and a temperature that was 4.8\% hotter than observations \citep{2014ApJ...797...31T}. The problems with the models are not unique to stars within binary systems. A study of 183 low-mass K7-M7 single stars by \citet{0004-637X-804-1-64} using an inferred stellar mass, found that models over-predicted the effective temperature ($T_{\rm eff}$) by 2.2\% and under-predict radii by 4.6\%. In some cases, magnetic fields have been used to explain the discrepancies, e.g. V530~Ori \citep{2014ApJ...797...31T}, but it remains unclear whether magnetic fields provide the solution in all cases \citep{0004-637X-804-1-64}. For low-mass stars, explanations for the discrepancies are discussed by \cite{2013AN....334....4T}, where they conclude that systems with well-determined masses, radii, temperatures and metallicities, will be important in trying to understand this problem.
 
AI Phoenicis (\object{AI Phe}, \object{HD 6980}) is one of a number of important eclipsing binary systems within astrophysics. Many other subgiant systems exhibit flares or spots that are associated with the strong magnetic activity of RS CVn systems. However, due its long period and slow rotation, AI Phe does not display any of these photometric complications \citep{1988A&A...196..128A}. This makes it possible to obtain high-precision masses and radii for the system, and treat the components as independent stars for modelling purposes. Its specific combination of a main-sequence star and a sub-giant star make it ideal for testing stellar evolutionary models. This was demonstrated by \citet{2010A&ARv..18...67T} using two different codes to model the system (Yonsei-Yale and an experimental version of the Victoria models). They aimed to reproduce the radii and effective temperatures at an age that was consistent for both components, and found the mean age of the system is different by 0.9\,Gyr between the two models. This is a non-negligible uncertainty when determining the ages of nearby solar-type stars. \citet{0004-637X-776-2-87} also used AI Phe to test their evolutionary code by attempting to produce tracks for each component that give consistent ages for both components. They found an age of $4.44\,$Gyr for the hotter component and $4.54$\,Gyr for the cooler component, with agreement at the 2\% level. The treatment of convective-core overshooting, mixing-length and helium abundance can be significant sources of uncertainty within the models \citep{2014EAS....65...99L}. With high precision masses and radii for a binary such as AI Phe, the age of the system can be tightly constrained, as the number of plausible evolutionary tracks is greatly reduced. It therefore provides an excellent opportunity to get a better understanding of overshooting, the helium abundance and mixing length parameter.

AI Phe was first noted as an eclipsing binary by \citet{1972IBVS..665....1S}. Photometric analysis to obtain the first accurate estimate of the orbital period was carried out by \citet{1978IBVS.1419....1R}. \citet{1979A&AS...36..453I} carried out the first spectroscopic analysis of the orbit. \citet{1984ApJ...282..748H} used the spectroscopic orbit data of \citet{1979A&AS...36..453I}, together with new photometric observations in {\em{UBVRI}}, to obtain masses and radii for AI Phe. \citet{1985ApJ...291..270V} compared the observed parameters of AI Phe to theoretical isochrones to calculate the helium abundance, $Y$, and age, $\tau$, of the system (\mbox{$Y=0.38\pm0.05$} and \mbox{$\tau=3.6\pm0.7$ Gyr}). \citet{1988A&A...196..128A} used new radial velocity (RV) measurements, the {\em{UBVRI}} photometric data of \citet{1984ApJ...282..748H} and new {\em{uvby}} photometry to obtain masses to $\pm$0.3\% and radii to $\pm$1.5\%. At the time, these were most accurately determined masses for any detached, double-lined eclipsing binary system \citep{1991A&ARv...3...91A}. \citet{1992ApJS...79..123M} remodelled the data of \citet{1988A&A...196..128A} and \citet{1984ApJ...282..748H}, using Wilson-Devinney code with updated model atmospheres. Their work reduced the uncertainties on the radii, although their value for the secondary component was 2-$\sigma$ lower than the value found by \citet{1988A&A...196..128A}. \citet{2007ChJAA...7..558K} reanalysed the RV data of \citet{1988A&A...196..128A} using a nonlinear regression method, which allowed semi-amplitude velocity, eccentricity and longitude of periastron to be fitted simultaneously. However, \cite{2009MNRAS.400..969H} suggested that the errors quoted by \citeauthor{2007ChJAA...7..558K} were underestimated due to several uncertainties not being included in their error analysis. \citet{2009MNRAS.400..969H} used new spectra to obtain RV measurements with a root mean squared (RMS) residuals from the spectroscopic orbit fit of 62 and 24 \mbox{m s$^{\rm -1}$} for the primary and secondary components, respectively. They also used photometric data from All-Sky Automated Survey \citep[ASAS, ][]{2002AcA....52..397P} to obtain masses and radii for the components, but noted that they only had access to one lightcurve for their work, and suggested the derived parameters of AI Phe could be improved further with high-precision photometry. 
	
	The SuperWASP cameras monitor thousands of stars every night looking for planetary transits, as part of the Wide Angle Search for Planets \citep[WASP,][]{2006PASP..118.1407P}. WASP also provides an excellent opportunity to find and study eclipsing binary systems using photometry with better than 1\% accuracy. This paper focuses on the analysis WASP photometry to derive high precision lightcurve parameters for AI Phe. These are then combined with the RV measurements of \citet{2009MNRAS.400..969H} to obtain radii to a precision of better than 1\%. Section~\ref{sec:data} provides a description of the photometric data used for the work. Section \ref{sec:Analysis} describes the data processing, lightcurve modelling, and error analysis. Section~\ref{sec:massAndRadii} contains a summary of the final masses and radii, while section~\ref{sec:models} uses the new masses and radii as constraints to determine the age of the system using models different mixing lengths and helium abundance. Finally, sections~\ref{sec:Discuss} and \ref{sec:Conc} contain a discussion of the results and conclusion, respectively.

\section{Data}
\label{sec:data}

\subsection{Observations} 
\label{subsec:Observations}

Located at the Observatorio del Roque de los Muchachos, La Palma and at Sutherland Observatory, South Africa, the two WASP instruments are both formed from eight wide-field cameras each with a 2048 x 2048 pixel CCD. For AI Phe (\object{1SWASP J010934.19-461556.0}), over 170\,000 photometric measurements were present in the WASP archive, taken between June 2006 and January 2014 by the instrument in South Africa. During this period, WASP-South has used two different types of lenses, the original 200-mm, f/1.8 lenses \citep{2006PASP..118.1407P} and, from July 2012,  85-mm, f/1.2 lenses \citep{2014CoSka..43..500S}. The reduction procedure is identical for both lens types, and uses a dedicated pipeline \citep{2006PASP..118.1407P} that has been optimised in each case. The data are then processed by a detrending algorithm, which was developed from the \textsc{SysRem} algorithm of \citet{2005MNRAS.356.1466T} and is described by \citet{2006MNRAS.373..799C}. 

Use of the 85-mm lenses allows WASP-South to focus on brighter stars ($V \lesssim9$). A larger reduction aperture is used in comparison to the 200-mm lenses (4 pixels from 3.5) and the limits on image and star rejection were also modified. This results in a shift of the photometry range to $6 \lesssim V \lesssim 11$ for the 85-mm lenses from $9\lesssim V \lesssim 13$ for the 200-mm lenses. The 200-mm lenses use broad-band filters with a range of 400-700\,nm, while the 85-mm use SDSS r' filters. Observations from the two different types of lens have been analysed separately.

\subsection{Initial Processing}
\label{subsec:InitialProcessing}

The WASP data can suffer from large amounts of scatter, due to clouds, etc. This section describes the methods used to remove unreliable observations in the two sets of data. 

For the 200-mm data, a decision was made to only use data from cameras 225 and 226 during the analysis. These two cameras contribute more than 80\% of the observations made with the 200-mm cameras. The data from the other cameras contained large amounts of scatter, perhaps related to the fact that the 200-mm lenses are very close to their saturation limit for AI Phe. They also only contributed a handful of observations to the eclipses, and would not help determine the radii of AI Phe. 

As part of the WASP reduction pipeline, each photometric measurement gets assigned a weighting factor, $\sigma_{\rm xs}$, \citep[denoted $\sigma_{t(i)}$ in ][]{2006MNRAS.373..799C}, to characterise the scatter present from external noise sources. This $\sigma_{\rm xs}$ value is set to zero if the pipeline deems the value to be missing or bad \citep{2006MNRAS.373..799C}, so data with $\sigma_{\rm xs}=0$ were not included in any analysis. In some cases, the error in the measured flux was exceptionally high in comparison with the other measurements. It is likely that this is due to clouds being present at the time of exposure, as $\sigma_{\rm xs}$ is also higher for these data. If the error in a particular flux measurement was five or more times greater than the median error of all data from the same camera, then it was excluded from our analysis. 

The flux measurement we have used for our analysis, $f$, is calculated from an aperture with a radius of 3.5 pixels or 4 pixels (for the 200-mm and 85-mm lenses, respectively) centred on AI Phe, with the detrending correction of \citet{2006MNRAS.373..799C} applied, and $f_{\rm err}$ is the associated error in this value. The magnitude of each measurement was calculated from $f$ using the median flux of the data set as the zero point. The error in magnitude, $m_{\rm err}$, associated with each observation, was calculated using 
\begin{equation}
\mbox{$m_{\rm err}$} = f \sqrt{\left(\frac{f_{\rm err}}{f}\right)^{2} + {\sigma_{\rm xs}}^2}
\label{eq:error}
\end{equation}

One final check was used to look for sections of data that were significantly offset from the remaining data. The phase-folded data was split into 800 phase bins. The median magnitude and associated error in the median were then calculated for each bin. This formed a model from which we could estimate the expected magnitude of an observation given its phase. The entire data set was also split into blocks based on the night the observation was taken and the camera used. For each data block, the magnitude of each observation within that block was compared to its expected magnitude by matching it to a phase bin. If more than 80\% of the observations from a particular data block were offset by more than ten times the errors associated with the appropriate phase-bins, then all data from that night/camera combination were excluded from the analysis.

Overall, from the initial 170\,324 observations stored in the WASP archive, 126\,780 remained for use during the analysis, 12\,618 from the 200-mm lenses and 114\,162 from the 85-mm lenses. Table~\ref{tab:AllPointsRemoved} provides a summary of the number of points removed during this initial processing stage for each type of lens.

\begin{table}
\caption{Number of observations removed during initial processing. The last row states the number of data remaining.}
\label{tab:AllPointsRemoved}
\centering
\begin{tabular}{l r r}
\hline\hline
\noalign{\smallskip}
Reason				& 200-mm 	& 85-mm \\
\noalign{\smallskip}
\hline
\noalign{\smallskip}
Removed cameras		& $3662$		& $0$		\\
$\sigma_{\rm xs} = 0$	& $2152$		& $16925$	\\
$f_{\rm err} >$ 5*median	& $1381$		& $18519$	\\
Offset				& $37$		& $868$		\\ 
\noalign{\smallskip}
\hline
\noalign{\smallskip}
Remaining			& $12618$	& $114162$	\\
\noalign{\smallskip} 
\hline
\end{tabular}
\end{table}

\section{Analysis}
\label{sec:Analysis}

\subsection{Ephemeris}
\label{subsec:ephemeris}

From an initial analysis of the data using the ephemeris given in \citet{1984ApJ...282..748H}, we found that the primary eclipse in the 85-mm data was offset in phase by $0.00102\pm0.00002$. This meant that the primary eclipse occurred more than 30 minutes later than predicted.

\begin{table}
\caption{Available times of primary minimum for AI Phe. For the 200-mm and 85-mm data, the days (in HJD$-2\,450\,000$) used to determine $t_{\rm pri}$ are included in the source column.}
\label{tab:TimeOfMin}
\centering
\begin{tabular}{l l l }
\hline\hline
\noalign{\smallskip}
$t_{\rm pri}$ (HJD$_{\rm UTC}$) 	& Error & Source \\
\noalign{\smallskip}
\hline
\noalign{\smallskip}
$2\,443\,410.6885$					& 0.0004		& {\citet{1978IBVS.1419....1R}}		\\
$2\,444\,861.6357$					& 0.0005		& {\citet{1984ApJ...282..748H}}		\\
$2\,453\,247.6306$					& 0.0027		& ASAS data					\\
$2\,454\,354.2869$					& 0.0016		& 200mm,	3890-4439		\\
$2\,455\,436.35626$					& 0.00013		& 200mm, 5370-5526		\\
$2\,455\,805.24418$					& 0.00014		& 200mm, 5739-5911		\\
$2\,456\,149.53828$					& 0.00012		& 85mm, 6111-6661			\\
\noalign{\smallskip} 
\hline
\end{tabular}
\end{table}

\begin{figure}
\resizebox{\hsize}{!}{\includegraphics{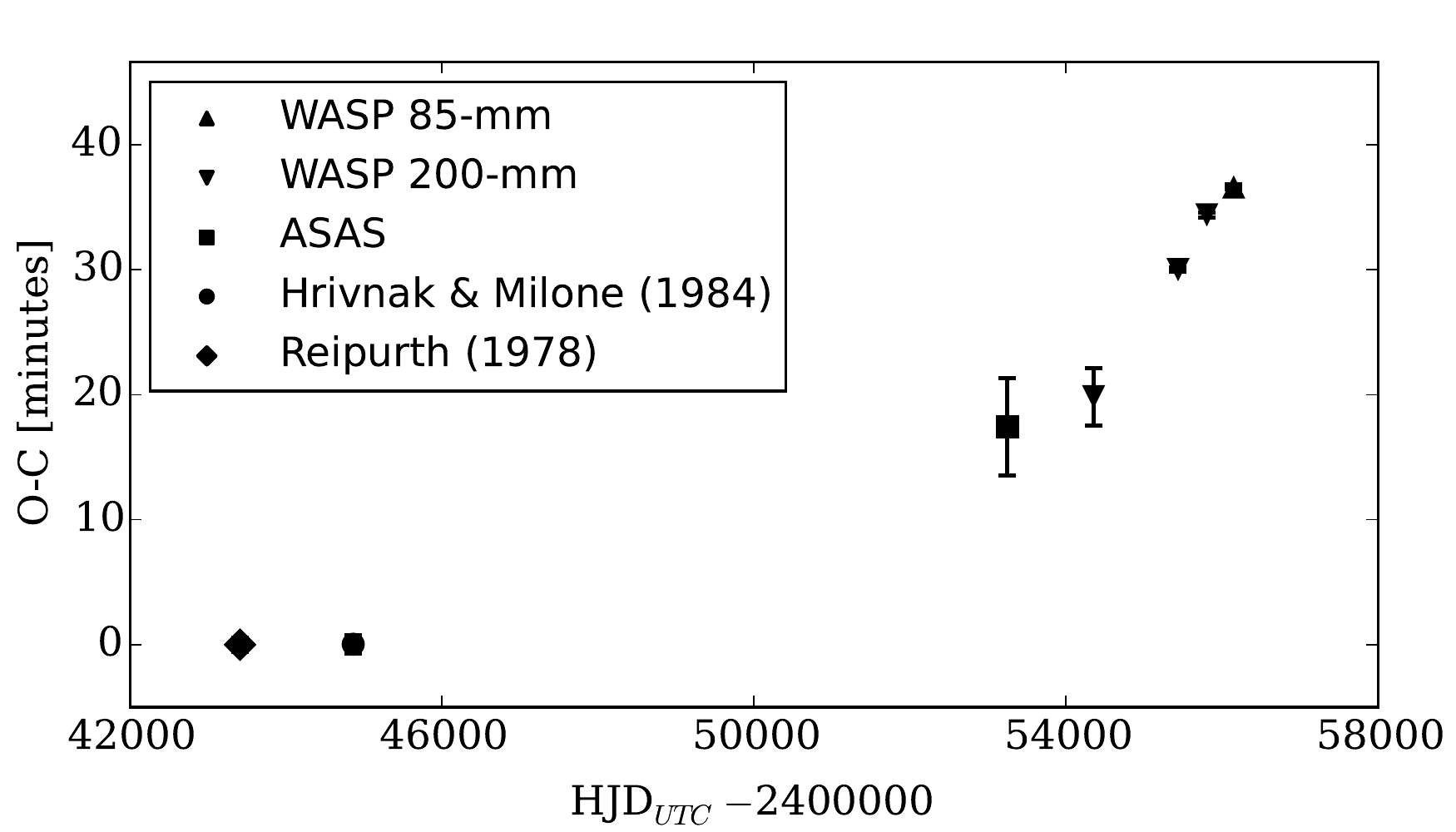}}
\caption{Comparison between observed and computed times of minima for the primary eclipse of AI Phe, using the ephemeris from {\citet{1984ApJ...282..748H}}.} 
\label{fig:tmin}
\end{figure}

Table~\ref{tab:TimeOfMin} details the times of primary minima, $t_{\rm pri}$, currently available for AI Phe. Two of the times have been previously published. The remaining five have been obtained by fitting data from WASP and All-Sky Automated Survey \citep[ASAS, ][]{2002AcA....52..397P} using \textsc{jktebop} \citep{JKTebopRef}. The 200-mm WASP data were split into three blocks -- the range of Heliocentric Julian Date used in each block is given in Table~\ref{tab:TimeOfMin}. Only one block was used from the 85-mm data. Figure~\ref{fig:tmin} shows the difference between observed $t_{\rm pri}$ and the calculated time based on the linear ephemeris of \citet{1984ApJ...282..748H}. Based on this plot, a linear ephemeris cannot be used to describe the long-term periodicity of AI Phe. There are an insufficient number of minima to obtain a reliable quadratic ephemeris. Therefore, for this work, the following linear ephemeris has been fitted to only times of minima for the WASP data:
\begin{equation}
\mbox{\rm HJD Pri. Min.} = 2\,455\,085.24370(21) + 24.592483(17)\,E.
\label{eq:ephemeris}
\end{equation}
For the shorter timescale covered by the WASP data, a linear ephemeris is a suitable approximation. The new period is consistent with the value determined spectroscopically by \citet{2009MNRAS.400..969H}, however, there is an 8.4-sigma difference between this new period and the value quoted in \citet{1984ApJ...282..748H}. The zero-point of the ephemeris from \citet{2009MNRAS.400..969H} was not included in Fig.~\ref{fig:tmin} and Table~\ref{tab:TimeOfMin}, because they use a different definition of this quantity. With the new ephemeris the phase-offset for the 85-mm data is reduced to 0.00005$\pm$0.00002.

\begin{figure}
\resizebox{\hsize}{!}{\includegraphics{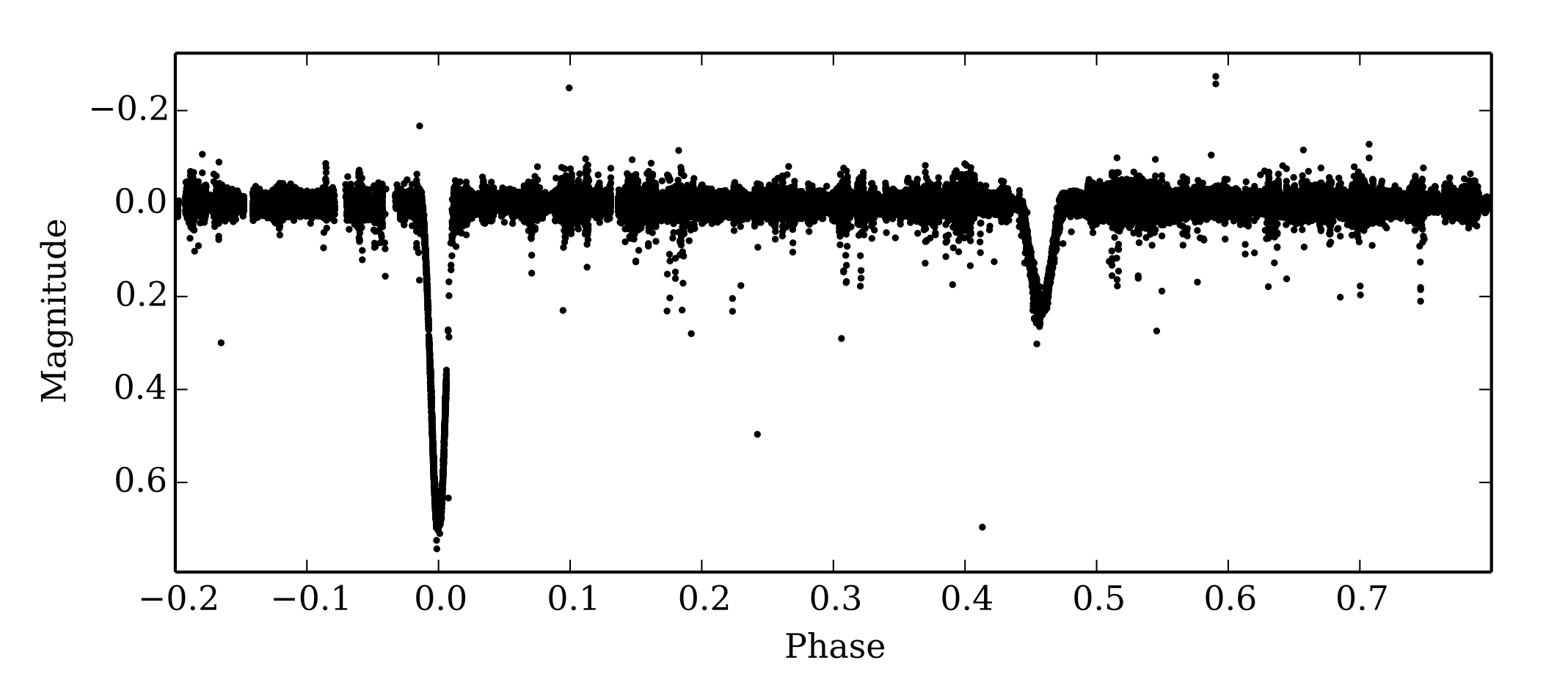}}
\resizebox{\hsize}{!}{\includegraphics{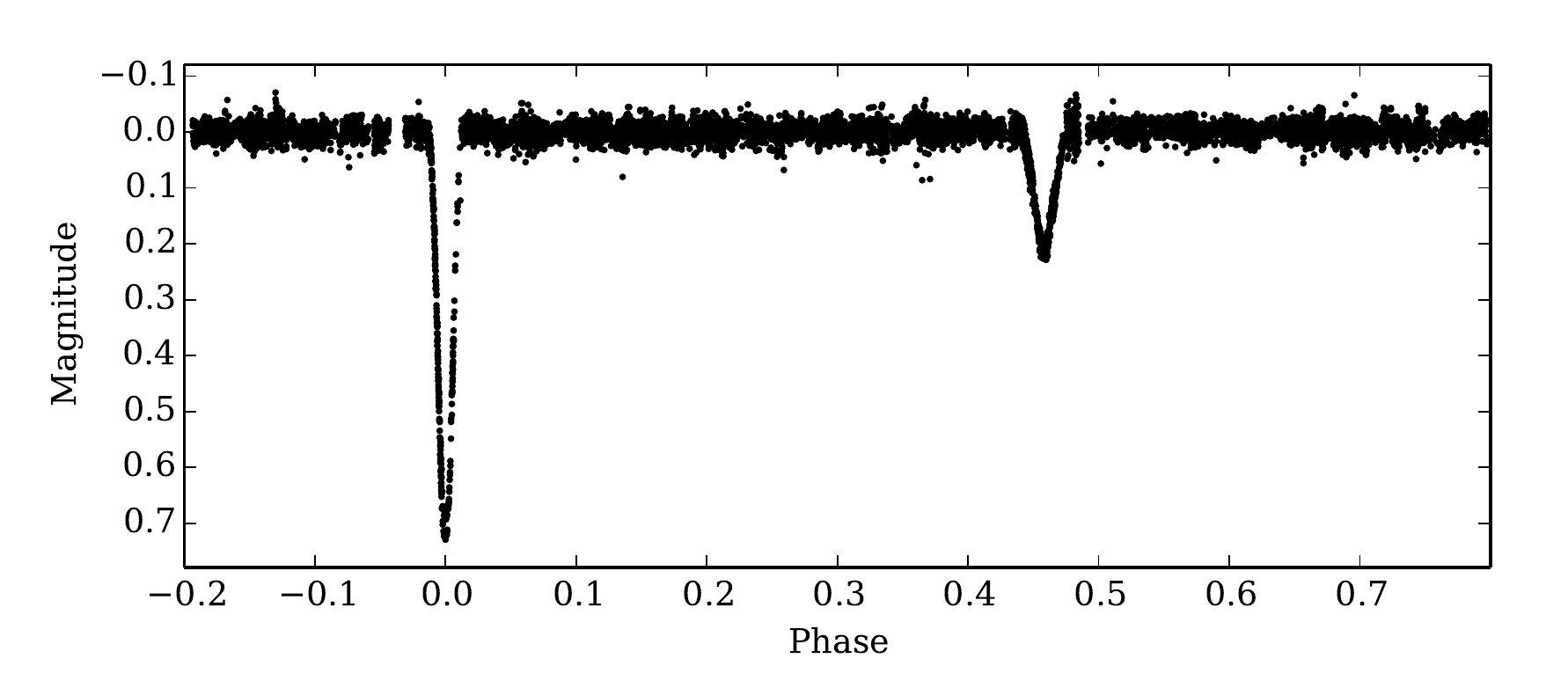}}
\caption{Top -- 85-mm WASP-South phase-folded lightcurve for AI Phe. Bottom -- Phase-folded WASP-South 200-mm data from cameras 225 and 226.} 
\label{fig:phaseFoldedLC}
\end{figure}

In all the observations of AI Phe over the last 40 years or so, the secondary eclipse has not been observed in its entirety in one night, so it has not been possible to investigate the timings of the secondary eclipse. This investigation would have given an insight into the possible cause of the apparently quadratic nature of the ephemeris. If the deviations in calculated timings are common to both the primary and secondary minima, it may suggest a third body is involved, or if the deviations have opposite signs then it may suggest apsidal motion is involved. 

Figure~\ref{fig:phaseFoldedLC} shows the phase-folded lightcurves for both the 85-mm and 200-mm data using the ephemeris in equation \ref{eq:ephemeris}, having been processed using the method described in Sect. \ref{subsec:InitialProcessing}.

\subsection{A Contaminating Companion}
\label{subsec:3rdlight}

\begin{figure}
\resizebox{\hsize}{!}{\includegraphics{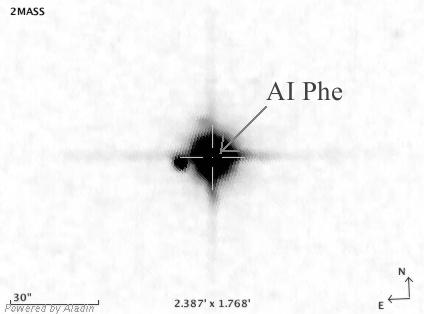}}
\caption{2MASS image (combined $J$, $H$ and $K_{\rm s}$ bands) of AI Phe showing the close proximity of the contaminating star \citep[Image: Aladin Sky Atlas, ][]{2000A&AS..143...33B}.} 
\label{fig:2MASSimage}
\end{figure}

An image of AI Phe from the Two Micron All Sky Survey \citep[2MASS, ][]{2006AJ....131.1163S} shows there is another star approximately 11\arcsec{} to the east of AI Phe, as shown in Fig. \ref{fig:2MASSimage}. The close proximity of the star means it resides within the aperture used for the WASP photometry of AI Phe and so contributes to the measured flux. As such, it has been necessary to include third light as a parameter that is fitted during the lightcurve analysis.

Although seen in the 2MASS images, this survey does not present any photometric measurements for this companion star. Near-infrared photometry for this star was taken from the Deep Near-Infrared Survey \citep[DENIS,][]{1997Msngr..87...27E}, with $I = 13.50\pm0.02$ mag, $J = 12.95\pm0.07$ mag and $K = 12.60\pm0.14$ mag. The $I-J$, $J-K$, and $I-K$ colours for the companion were compared to the Dartmouth stellar evolution models \citep{2008ApJS..178...89D}, and by assuming a main-sequence star with $[{\rm Fe}/{\rm H}]=0.0$ and an age 2 Gyr, its mass was estimated as $0.91\pm0.06 M_{\sun}$. This was used to obtain expected absolute magnitudes from the Dartmouth model, which were in turn used to calculate the distance modulus in each band, resulting in an estimated distance of \mbox{$590\pm9$pc} to the companion. Comparing this value with the distance to AI Phe of \mbox{$162\pm6\,{\rm pc}$} \citep{1988A&A...196..128A}, we conclude that the visual companion is not physically associated with the system.

\subsection{Model}
\label{subsec:Model}

This work uses the \textsc{ebop} lightcurve analysis code \citep{1972ApJ...174..617N, 1981AJ.....86..102P}. The subroutine {\tt{light}} from this code, which calculates the lightcurves at specified orbital phases for a given set of model parameters, was converted to double-precision floating point arithmetic and modifications were made to enable it to be called directly from the programming language, Python. The \textsc{ebop} lightcurve is then combined with the least-squares Levenberg-Marquardt, Python module, \textsc{MPFIT} \citep{2009ASPC..411..251M} to find the best-fit parameters, i.e. the parameters that minimise the $\chi^2$ value between the model and data. 

The following seven parameters were allowed to vary during the fitting process: surface brightness ratio at the centre of the stellar discs, $J$; sum of the radii, $r_{\rm sum} = r_{1} + r_{2}$; ratio of the radii, $k = r_{2}/r_{1}$; inclination, $i$; $e\cos{\omega}$, $e\sin{\omega}$; third-light, $l_{3}$. In addition to these fitted parameters, the fractional radii, $r_{1}$ and $r_{2}$ are automatically calculated from $r_{\rm sum}$ and $k$, while the eccentricity $e$ and longitude of periastron $\omega$ are calculated from $e\cos{\omega}$ and $e\sin{\omega}$. 

The mass ratio, $q = M_{2}/M_{1}$, was fixed at 1.0418, taken from \citet{2009MNRAS.400..969H}. This is slightly larger than the value 1.034 quoted by \citet{1988A&A...196..128A}, but modifying the mass ratio by such a small amount had no effect on the best-fit parameters. The gravity darkening exponents also have little impact on the shape of the lightcurve in the case of AI Phe as the system is well-detached. Using values taken from the table by \citet{2011A&A...529A..75C}, the gravity darkening exponent for the primary, $y_{\rm p}$, and secondary, $y_{\rm s}$, were fixed at 0.26 and 0.50 respectively. The same values were used for both the 85-mm and 200-mm data.

The linear limb-darkening coefficients were also held fixed. Attempts were made to include these as free parameters, but we found that they are not usefully constrained by the data. Instead, the limb darkening coefficients were estimated by interpolation in the table of \citet{2011A&A...529A..75C} by using values for effective temperature, surface gravity and metallicity for the two components from \citet{1988A&A...196..128A}. For the 85-mm data we adopted the primary limb darkening coefficient, $u_{\rm p}=0.54\pm0.03$ and $u_{\rm s}=0.67\pm0.03$ for the secondary. For the 200-mm data we used $u_{\rm p}=0.52\pm0.05$ and $u_{\rm s}=0.67\pm0.05$ using the {\em Kepler} pass-band to approximate the response of the WASP broad-band filter. To account for the uncertainties in the limb darkening coefficients, each coefficient was varied by its error and a new model fitted. For each parameter, the average absolute difference between these models and the best fit model has been added in quadrature to the uncertainties from the best-fit parameters. Table~\ref{tab:LDuncert} details the typical contribution to the uncertainty of each parameter, for both the 200-mm and 85-mm data. These were calculated from the mean contributions for each best-fit detailed in Sects.~\ref{subsec:Detrend} and \ref{subsec:constraints}.

\begin{table}
\caption{Typical uncertainty contribution to each parameter from uncertainty limb darkening coefficients.}
\label{tab:LDuncert}
\centering
\begin{tabular}{l r r}
\hline\hline
\noalign{\smallskip}
Parameter				& 200-mm 	& 85-mm \\
\noalign{\smallskip}
\hline
\noalign{\smallskip}
$J$				& $0.0066$	& $0.0051$	 \\
$r_{\rm sum}$		& $0.00015$	& $0.00010$	   \\
$k$ 				& $0.006$	 	& $0.004$		   \\
$i$ (\degr)			& $0.062$		& $0.008$		   \\
$e \cos \omega$ 	& $0.00002$ 	& $0.00002$	  \\
$e \sin \omega$ 	& $0.0013$ 	& $0.0011$	  \\
$l_{3}$			& $0.004$		& $0.003$		  \\
\noalign{\smallskip}
\hline
\noalign{\smallskip}
$r_1$ 			& $0.00018$ 	& $0.00010$ 	  \\
$r_2$			& $0.000013$	& $0.00009$	   \\
$e$				& $0.0012$	& $0.0010$		\\
$\omega$ (\degr)	& $0.12$		& $0.13$		 \\
\noalign{\smallskip} 
\hline
\end{tabular}
\end{table}

\subsection{Parameter-Space Exploration with \texttt{emcee}}
\label{subsec:EMCEE}

One of the greatest risks when using a least squares minimisation method is that the solution that is found is a local minimum, rather than the overall global minimum of the problem. If local minima are present it is important that these are considered when the uncertainties for the final parameters are calculated. Markov Chain Monte Carlo (MCMC), in the form of the Python module, \texttt{emcee}, \citep{GoodmanWeare2010} was used to check for these local minima, within the parameter-space of the same seven parameters used in the model fitting ($J$, $r_{\rm sum}$, $k$, $i$, $e\cos{\omega}$, $e\sin{\omega}$ and $l_3$). \texttt{emcee} uses the affine-invariant ensemble sampling (stretch-move) algorithm developed by \cite{GoodmanWeare2010}, where a group of walkers explore the parameter space. This group of walkers can be split allowing the process to be run in parallel and the affine-invariant transformations mean the algorithm can cope with skewed probability distributions. The positions of the walkers within a particular sub-group are updated using the positions of walkers in the other subgroups. 

The probability that a model produced by the parameters from the walkers, corresponds to the best-fit model is evaluated using the log likelihood function
\begin{equation}
\ln \mathcal{L}({\bf y }; {\bf \Theta}) = -\frac{1}{2} \sum^N_{n=1} \left[ \left(\frac{m_{n}-y_{n}({\bf \Theta})}{m_{{\rm {err,}}n}}\right)^2 - \ln \left( \frac{2\pi}{m_{{\rm{ err,}}n}^2}\right) \right]
\label{eq:loglike}
\end{equation}
where {\bf y} is a vector of length $N$ containing the magnitudes generated for a model, {\bf $\Theta$} is a vector containing the varying parameters ($J$, $r_{\rm sum}$, $k$, $i$, $e\cos{\omega}$, $e\sin{\omega}$ and $l_3$), $m$ is the observed magnitude and $m_{\rm err}$ is the standard error on the magnitude. Priors were applied, but these are only used to prevent the parameter exploring areas that are unphysical, eg. $r_{\rm sum}$ or $J$ being less than zero. 

The process ran using 150 walkers for 2\,500 steps, of which the first 200 were discarded to allow for an adequate burn-in stage. For each of the walkers, a starting point for each parameter was chosen by choosing a number at random from a normal distribution (with a mean of zero and variance of 0.01) and adding it to the best-fit parameter. While this method would initially create a ball of walkers close to the solution found by the model, the burn-in stage allows the walker to spread out from this ball. Each parameter was subsequently checked, by plotting the walkers' positions against step number, to ensure the burn-in stage was completed within these first 200 steps. To ensure the was no bias from the walkers' starting positions, a test was run that allowed the walkers to start further from the model solution, up to three times the parameters' uncertainties as determined from the covariance matrix of the best-fitting model. The burn-in stage for this test was much longer, but also failed to reveal any other local minima, so we concluded the original starting points were adequate.

\begin{figure*}
\centering
\resizebox{\hsize}{!}
{\includegraphics[width=18cm]{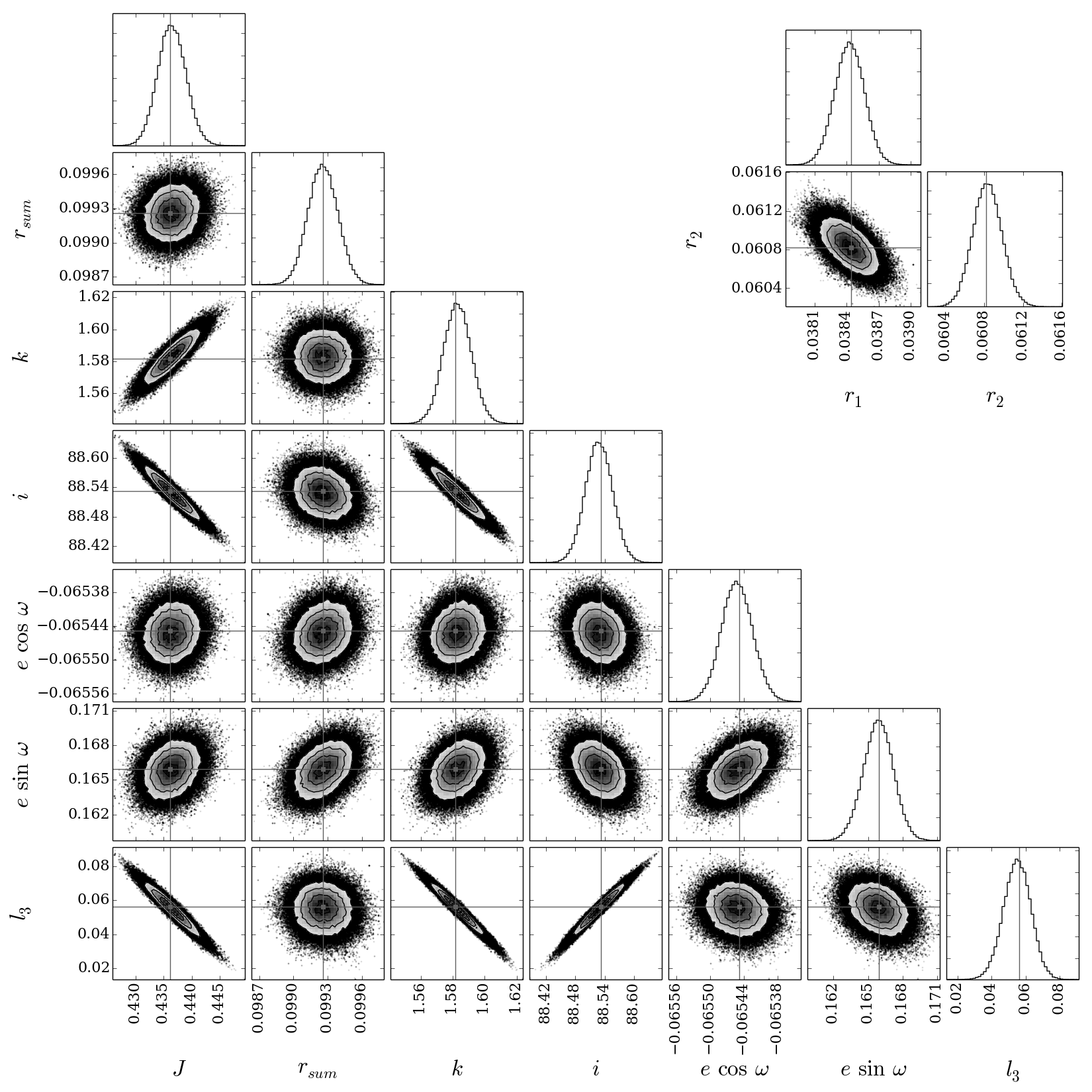}}
\caption{Density distribution of parameter space explored using MCMC. The best-fit parameters as determined by the model are marked by the grey crosses on the density distributions, and the grey line on the histograms. The contours on each of the distributions indicate the density of the points, with the darker, denser regions towards the centre of each plot. The distribution for $r_{1}$ and $r_{2}$ has been calculated from $k$ and $r_{\rm sum}$. Data shown is for the 85-mm original data, without priors. } 
\label{fig:MCMCdetrend}
\end{figure*}

Figure~\ref{fig:MCMCdetrend} shows example distributions for the 85-mm data. The contours plotted over the top of the distributions indicate the density of points with the darker regions showing higher densities towards the centre of each plot. The grey lines across each of the distributions show where the best-fit values lie. Generally the resulting distributions are symmetric, as shown by the histograms for each parameter. However, there are correlations between some of the parameters, as seen by the tilted ellipsoid shaped regions. Many of these correlations link the third-light parameter to the other parameters, more specifically there are strong correlations between $l_{3}$ and $J$, $k$ and $i$. The correlation with $k$ highlights the importance for the inclusion of the third-light parameter in the model, because without it, the resulting relative radii would be subject to a systematic error. 

\subsection{Prayer-Bead Error Analysis}
\label{subsec:Prayer-bead}

It is inappropriate to use the covariance matrix from the least-squares fit to estimate the standard errors on the model parameters. This is because the assumption of uncorrelated noise with a Gaussian distribution is not satisfied for WASP photometry. It is not clear that the distribution of points from the MCMC chain gives an accurate impression of the posterior probability distribution of the parameters, for the same reason. Therefore, it is not appropriate to use the likelihood calculated using equation \ref{eq:loglike} either. As such, the standard errors on the best-fit parameters have been calculated using a prayer-bead (residual permutation) as described by \citet{2008MNRAS.386.1644S} and based on an algorithm developed by \citet{2002ApJ...564..495J}. The residuals between the data and the best-fit model are shifted by a number of steps to create a synthetic data set.  A new model is then fitted to this synthetic data, and the process is repeated across the entirety of the original data set. The uncertainties are the standard deviation of the fitted parameters from all the synthetic data models. Ideally, the number of shifts would be $N-1$, where $N$ is the number of observations used in generating the best-fit model. However, due to the large number of points involved in the WASP lightcurves, the number of shifts used was restricted to 500 spread evenly across the data. To ensure the best-fit solution was not affected by the choice of initial parameters, the initial parameters were taken at random from positions within the MCMC analysis.

\subsection{Detrending Investigations}
\label{subsec:Detrend}

\begin{table*}
\caption{Best-fit parameters for AI Phe from 85-mm and 200-mm data, with and without the detrending applied. Standard errors on the final two digits of each parameter value are given in the parentheses and include the contribution from the uncertainties in the limb darkening coefficients used.}
\label{tab:DetrendData}
\centering
\begin{tabular}{l r r r r r r}
\hline\hline
\noalign{\smallskip}
Parameter			& \multicolumn{3}{c}{85-mm}			& \multicolumn{3}{c}{200-mm}	 \\
				& Original		& Detrended	 & Difference	& Original 		& Detrended	 & Difference\\
\noalign{\smallskip}
\hline
\noalign{\smallskip}
$J$				& $0.4361(78)$		& $0.4336(67)$		& $0.0025$	& $0.367(15)$		& $0.391(17)$	& $-0.024$	 \\
$r_{\rm sum}$		& $0.09926(41)$	& $0.09902(32)$	& $0.00024$	& $0.1007(11)$		& $0.1003(9)$	& $0.0004$   \\
$k$ 				& $1.582(18)$	 	& $1.578(14)$		& $0.004$ 	& $1.512(51)$ 		& $1.558(50)$	& $-0.046$     \\
$i$ (\degr)			& $88.531(60)$		& $88.549(48)$		& $-0.018$	& $88.76(19)$ 		& $88.62(18)$	& $0.14$    \\
$e \cos \omega$ 	& $-0.06545(10)$ 	& $-0.06559(7)$	& $0.00014$	& $-0.06455(31)$ 	& $-0.06457(25)$	& $0.00002$	  \\
$e \sin \omega$ 	& $0.1659(40)$ 	& $0.1636(33)$		& $0.0023$ 	& $0.194(11)$		& $0.192(9)$	 & $0.002$	  \\
$l_{3}$			& $0.056(17)$		& $0.059(13)$		& $-0.003$	& $0.145(39)$ 		& $0.098(42)$	& $0.047$	  \\
\noalign{\smallskip}
\hline
\noalign{\smallskip}
$r_1$ 			& $0.03844(46)$ 	& $0.03841(37)$ 	& $0.00003$	& $0.0399(14)$ 	& $0.0392(13)$ & $0.0007$  \\
$r_2$			& $0.06082(38)$	& $0.06061(29)$	& $0.00021$	& $0.0607(11)$		& $0.0611(10)$	& $-0.0004$   \\
$e$				& $0.1784(43)$ 	& $0.1763(35)$		& $0.0021$	& $0.205(12)$		& $0.203(10)$	& $0.002$\\
$\omega$ (\degr)	& $111.52(45)$ 	& $111.83(39)$		& $-0.31$		& $108.39(98)$		& $108.59(80)$	 & $-0.20$ \\
\noalign{\smallskip}
\hline
\end{tabular}

\end{table*}

As mentioned in Sec. \ref{subsec:Observations}, WASP photometry is processed using the detrending algorithm described in \citet{2006MNRAS.373..799C}. The algorithm is used to remove four trends of systematic errors which are common to all stars in a particular field and is given by the equation
\begin{equation}
\widetilde{m}_{i, j} = m_{i,j} - \sum^{M}_{k=1} {^{(k)}}c_{j} {^{(k)}}a_{i}
\label{eq:detrendCC}
\end{equation}
where $m_{i,j}$ and $\widetilde{m}_{i,j}$ are the observed and corrected magnitude, respectively, for star $j$ at time $i$. M is the total number of trends, $a_{i}$ are basis functions detailing the patterns of systematic errors and $c_{j}$ describes to what extent each basis function affects a particular star. The algorithm is applied separately to each unique combination of camera and season, and aids the process locating planetary transits by flattening lightcurves over the time scale of a typical transit length (2.5 hours). The work in this paper aims to determine the radii of AI Phe to the highest accuracy possible using the WASP photometry, and so the effects of the detrending process on the resulting parameters have been investigated.

Reversing the detrending entirely would re-introduce all the systematic errors the algorithm was designed to remove, making comparisons between the resulting parameters more difficult. Instead, effective detrending coefficients, $c{'}$, have been calculated. These coefficients take into consideration the variability of an eclipsing binary, by including a lightcurve model, $L$, when they are calculated. For AI Phe, the situation can be described using 
\begin{equation}
\widetilde{m}_{i} = m_{i} - \sum^{M}_{k=1} {^{(k)}}c{'} {^{(k)}}a_{i} + L_{i}
\label{eq:detrend}
\end{equation}
where $m_{i}$ and $\widetilde{m}_{i}$ are the observed and corrected magnitude (respectively) for AI Phe, and $ {^{(k)}}a_{i}$ are the same detrending basis functions as before. The effective detrending coefficients for each of the basis functions were calculated using singular value decomposition (SVD). Effects from the altered detrending coefficients were removed from the observed data, then using \textsc{ebop} and \textsc{\mbox{MPFIT}}, best-fit parameters were obtained. A best-fit model generated {\bf{using the method}} in Sec. \ref{subsec:Model}, was used for $L$ initially. However, to ensure the choice of initial model did not bias the results, the values of $c{'}$ were calculated for multiple models. This was done as an iterative process, which continued until all parameters change by less than 0.005\% with each new model. Once the final set of best-fit parameters was determined, MCMC and prayer-bead analysis were used to calculate their associated uncertainties.

 \begin{figure}
\centering
\resizebox{\hsize}{!}{\includegraphics{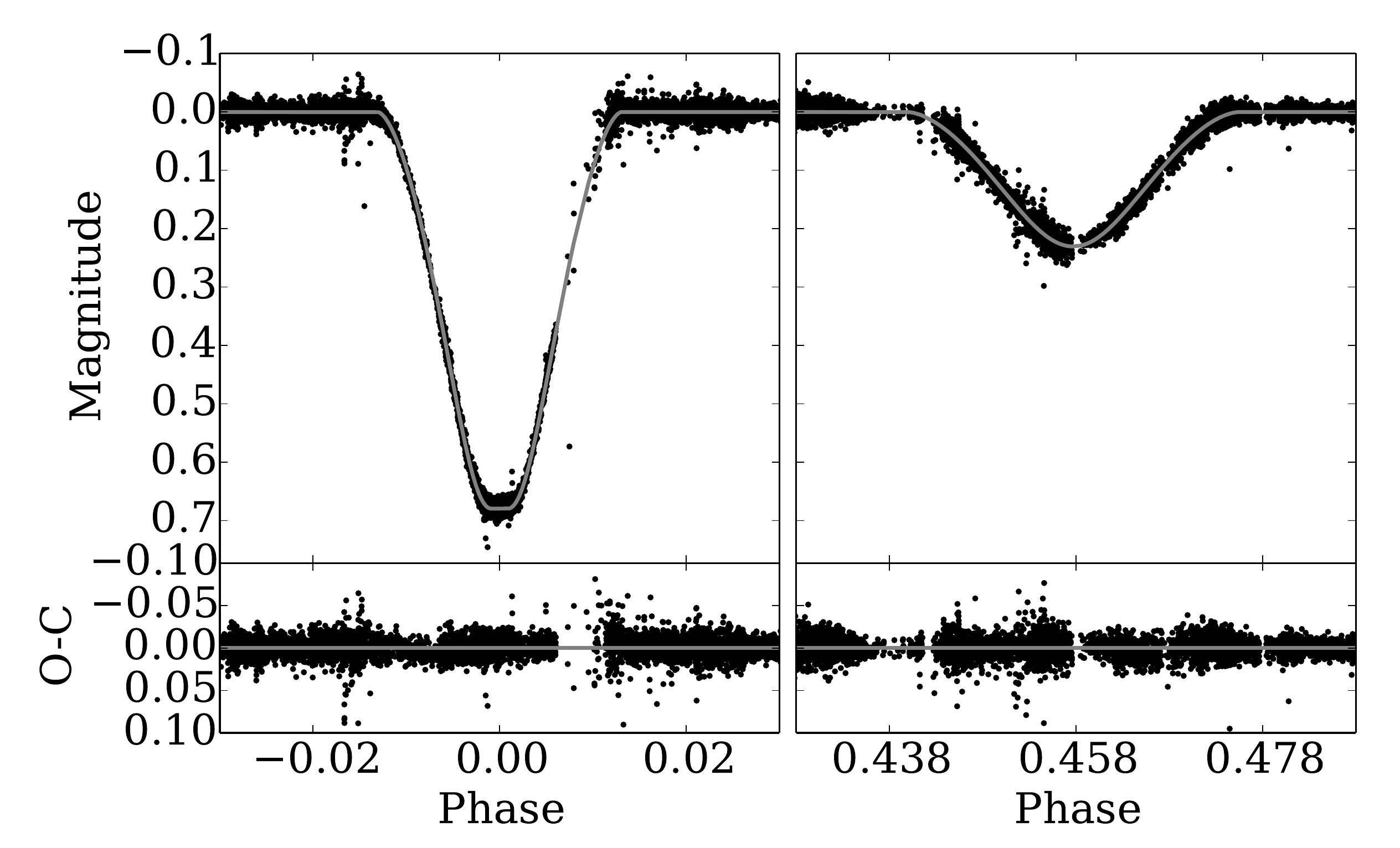}}
%\resizebox{\hsize}{!}{\includegraphics{eclipse85mmTrend2.eps}}
\caption{In the upper panels, the detrended best-fit model for AI Phe (grey line) plotted over the 85-mm WASP-South photometry for the primary (left) and secondary (right) eclipses. In the lower panels are the residuals between the plotted model and the data, with the grey line marking zero.} 
\label{fig:bestFit85}

\resizebox{\hsize}{!}{\includegraphics{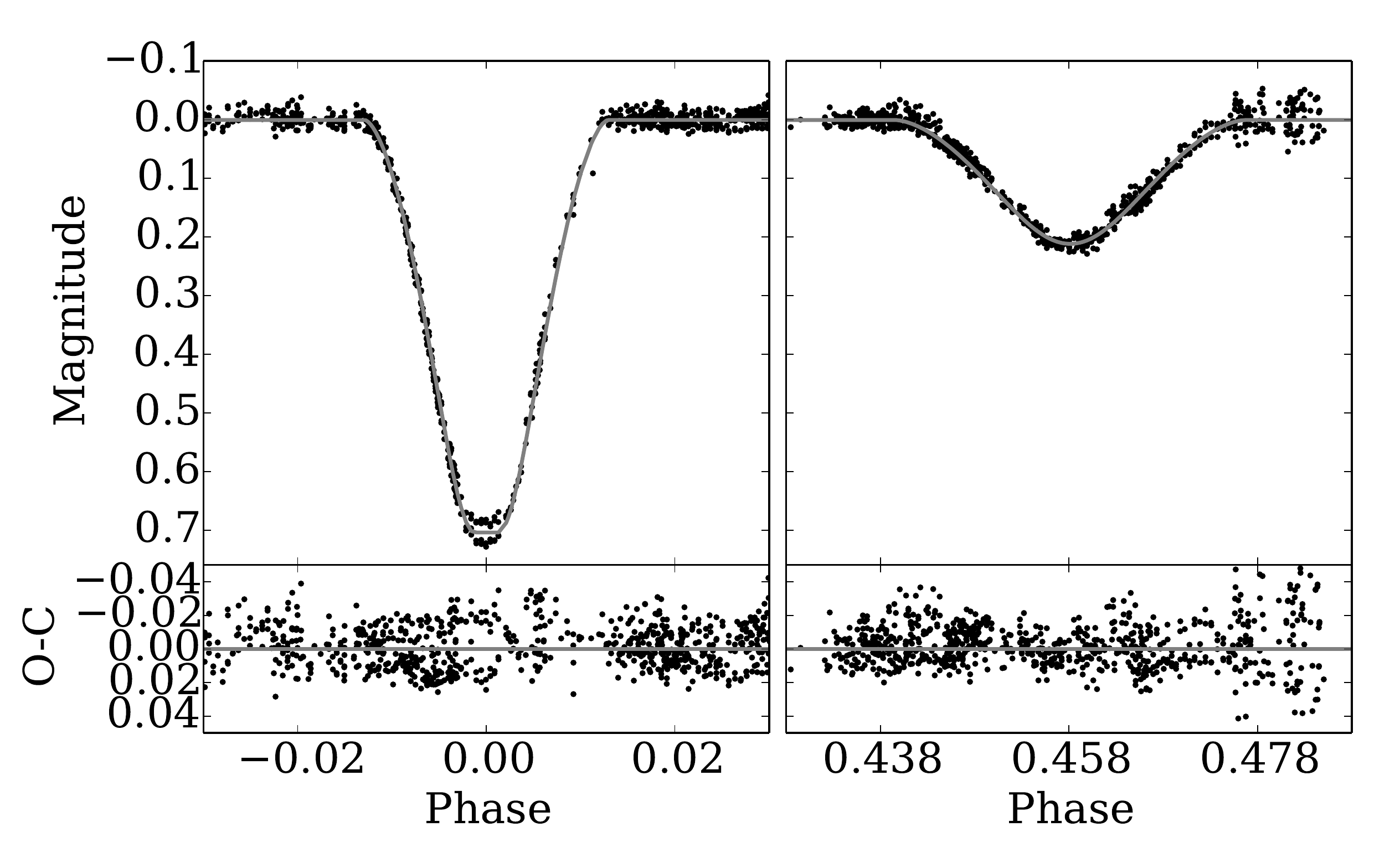}}
\caption{In the upper panels, the best-fit model for AI Phe (grey line) plotted over the 200-mm WASP-South photometry for the primary (left) and secondary (right) eclipses. In the lower panels are the residuals between the plotted model and the data, with the grey line marking zero.} 
\label{fig:bestFit200}

\end{figure}

Table \ref{tab:DetrendData} contains the best-fit parameters for both the 85-mm and 200-mm, with the difference between the original data and detrended data included for comparison. With the exceptions of $J$ and $l_3$ for the 200-mm data and $e\cos \omega$ for the 85-mm, all the best-fit parameters for the detrended data lie within the uncertainties of the parameters from the original data. Therefore, in general, the detrending algorithm applied during the WASP pipeline does not affect the overall best-fit parameters. The large difference present for $J$ and $l_3$ for the 200-mm data is due to the quality of the data present in the primary eclipse. The eclipse is covered by the two cameras on one night, and there is an offset between the data from the two cameras. This maybe be due to the differences in the transmission profile of the filters used in different cameras. In the detrended case, the model shifts enough to favour the data from camera 226, which forms a deeper eclipse. This results in the larger value of $J$ and, because the surface brightness is strongly correlated with $l_3$, there is the corresponding decrease in $l_3$. Figures \ref{fig:bestFit85} and \ref{fig:bestFit200} show the best-fit model plotted against the detrended data for the 85-mm and 200-mm data, respectively. The fit residuals are also shown in these figures.

As the sets of parameters from the original and detrended data are generally consistent, but the uncertainties associated with the detrended parameters are smaller, it is the detrended parameters that will be used from now on.

\subsection{Constraining $e\cos \omega$ and $e\sin \omega$}
\label{subsec:constraints}

\begin{table*}
\caption{Best-fit parameters for AI Phe from detrended 85-mm and 200-mm data, with the priors, $e \cos \omega =-0.064\pm0.004$ and $e \sin \omega =0.176\pm0.003$. The difference from the best-fit parameters without priors is included for comparison. Standard errors on the final two digits of each parameter value are given in the parentheses and include the contribution from the uncertainties in the limb darkening coefficients used. }
\label{tab:PriorData}
\centering
\begin{tabular}{l r r r r}
\hline\hline
\noalign{\smallskip}
Parameter		& \multicolumn{2}{c}{85-mm}							& \multicolumn{2}{c}{200-mm}	 \\
			& \multicolumn{1}{c}{Detrend}	& \multicolumn{1}{c}{Difference} & \multicolumn{1}{c}{Detrend}	& \multicolumn{1}{c}{Difference} \\
			& \multicolumn{1}{c}{with priors}& \multicolumn{1}{c}{to no priors}& \multicolumn{1}{c}{with priors}& \multicolumn{1}{c}{to no priors}	\\
\noalign{\smallskip}
\hline
\noalign{\smallskip}
$J$				& $0.4346(68)$		& $-0.0010$		& $0.388(15)$		& $0.003$		\\
$r_{\rm sum}$		& $0.09909(31)$	& $-0.00007$		& $0.0997(7)$ 		& $0.006$  	\\
$k$ 				& $1.582(15)$		& $-0.004$		& $1.546(47)$ 	 	& $0.012$ 	\\
$i$ (\degr)			& $88.535(48)$		& $0.014$			& $88.68(17)$  		& $-0.06$		\\
$e \cos \omega$ 	& $-0.06558(7)$	& $0.00001$ 		& $-0.06468(21)$	& $0.00011$ 	\\
$e \sin \omega$ 	& $0.1654(28)$		& $-0.0018$ 		& $0.185(5)$		& $0.007$ 	\\
$l_{3}$			& $0.057(13)$		& $0.002$			& $0.107(39)$		& $0.009$		\\
\noalign{\smallskip}
\hline
\noalign{\smallskip}
$r_1$ 			& $0.03838(37)$ 	& $0.00003$ 		& $0.0392(12)$ 	& $0.0000$ 	\\
$r_2$			& $0.06071(29)$	& $-0.00010$ 		& $0.0606(8)$ 		& $0.0005$	\\
$e$				& $0.1780(30)$		& $0.0017$ 		& $0.196(6)$		& $0.007$ 	\\
$\omega$ (\degr)	& $111.63(32)$		& $0.20$ 			& $109.28(46)$	 	& $-0.69$		\\
\noalign{\smallskip}
\hline

\end{tabular}
\end{table*}

In Table \ref{tab:DetrendData}, between the 85-mm and 200-mm parameters, there are differences of $2.7\sigma$ and $4\sigma$ for $e$ and $\omega$, respectively. The values also differ from the values quoted by \citet{2009MNRAS.400..969H} and \citet{1988A&A...196..128A}, $e$: 0.187(4) and 0.188(2) respectively; $\omega$: 110.1(9)\degr{} and 109.9(6)\degr{}, respectively. To investigate further, $e\cos \omega$ and $e\sin \omega$ were fixed at $-0.06424$ and $0.17561$, respectively. These values were calculated from the spectroscopic $e$ and $\omega$ of \citet{2009MNRAS.400..969H} and were chosen over \citet{1988A&A...196..128A}, as the \citeauthor{2009MNRAS.400..969H} values were obtained more recently. If the orbit has been varying (see Sec. \ref{subsec:ephemeris}), the most recent values should be nearest to those applicable to the time span covered by the WASP data.

However, fixing $e\cos \omega$ and $e\sin \omega$ produced models that were poor fits to the data. This was the case for both the 85-mm and the 200-mm data, but was more significant in for the 85-mm. For the 85-mm data, there was a phase offset of 0.001 between the observed data and resulting model of the secondary eclipse. 

As an alternative to fixing $e\cos \omega$ and $e\sin \omega$, the values calculated from \citet{2009MNRAS.400..969H} with their standard errors were used as Gaussian priors during the model fitting. Table \ref{tab:PriorData} contains the best-fit parameters for the detrended 85-mm and 200-mm data, with inclusion of the priors. Again, the uncertainties have been calculated through MCMC and prayer-bead analysis, and the error contribution from the uncertainties in the limb darkening value has also been included. For the 85-mm data, the inclusion of the priors has altered the best-fit parameters by less than their uncertainties. $e$ and $\omega$ have been altered more significantly for the 200-mm data, bringing their values closer to those of the 85-mm data. However, the values for both are still inconsistent with each other.

The exact cause of the differing $e$ and $\omega$ values remains unclear.  Previous observations of AI Phe have yielded a range of values for $e$ and $\omega$. For example, \cite{1984ApJ...282..748H} found $e = 0.1726\pm0.0006$ and $\omega= (111.8\pm0.1)\degr$ giving $e\cos \omega = -0.06410\pm0.00007$ and $e\sin \omega=0.1603\pm0.0007$, while the same {\em{UBVRI}} lightcurve analysed by \citet{1988A&A...196..128A} yielded mean values of $e\cos \omega = -0.064$ and $e\sin \omega=0.183$. No clear trend is present when all available value of $\omega$ were plotted against time, as might be expected if these differences are due to apsidal motion. The parameter $e\sin \omega$ is very sensitive to the shape of the secondary eclipse. Without observing the base of the secondary eclipse in one night, defining the exact shape of the eclipse and therefore determining the value of $e\sin \omega$ can be difficult. Values of $e$ and $\omega$ determined from spectroscopic orbits will not suffer these problems and should therefore be more accurate. It seems there is still work to be done in order to completely understand the behaviour of AI Phe's orbit.

Despite the inconsistency of $e$ and $\omega$, the addition of the priors has had very little impact on $r_{1}$ and $r_{2}$. They have remained consistent with each other, with a small reduction in the uncertainties of the 200-mm data. The radii are determined by the contact points, which are well defined by the primary eclipse, and the ratio of the eclipses $k$, but only as $k^{0.25}$. Therefore, $r_{1}$ and $r_{2}$ are robustly measured despite problems with the secondary eclipse and small changes in the 200-mm eclipse depth. A number of the other parameters have also shown small reductions in their uncertainties, and therefore the best-fit parameters obtained with the priors have been used in further analysis.

\subsection{Overall best-fit parameters}
\label{subsec:best fit params}

Table~\ref{tab:AIPheParamData} summarises the results from the lightcurve analysis, and includes the sets of best-fit parameters from the detrended 85-mm and 200-mm data, obtained with priors. For comparison, the results of \citet{1988A&A...196..128A} are also included in the table.  As the two WASP data sets and the results of \citet{1988A&A...196..128A} are all independent, they have been combined to produce an overall weighted mean. Uncertainties from each of the parameters were used to calculate internal and external standard errors on these mean values (i.e., based on the error bars and scatter, respectively), with the larger of the two being quoted alongside the weighted means. For $r_1$ and $r_2$ the results were internally consistent. The mean surface brightness ratio and third-light are not included because the different filters were used to obtain the 200-mm and 85-mm data.

\begin{table*}
\caption{AI Phe lightcurve parameters summary from this study, and comparative results from {\citet{1988A&A...196..128A}}.}
\label{tab:AIPheParamData}
\centering
\begin{tabular}{l r r r r }
\hline\hline
\noalign{\smallskip}
Parameter					& 85-mm		& 200-mm		& {\citet{1988A&A...196..128A}} 	& Weighted mean \\
\noalign{\smallskip}
\hline
\noalign{\smallskip}
Surface brightness ratio, $J$	& $0.4346(68)$		& $0.388(15)$		& --					& -- \\
Sum of radii, $r_{\rm sum}$	& $0.09909(31)$	& $0.0997(7)$		& 0.0993(10)\parbox{0pt}{$^*$}	& $0.09919(27)$   \\ 
Ratio of radii, $k$ 			& $1.582(15)$	 	& $1.546(47)$		& $1.613(10)$ 			&  $1.602(12)$    \\
Inclination, $i$ (\degr)		& $88.535(48)$		& $88.68(17)$		& $88.45(5)$ 			& $88.502(39)$    \\ 
$e \cos \omega$ 			& $-0.06558(7)$ 	& $-0.06468(21)$	& $-0.0634(3)$ 			& $-0.06534(40)$  \\ 
$e \sin \omega$ 			& $0.1654(28)$ 	& $0.185(5)$		& $0.178(10)$ 			& $0.1710(61)$  \\ 
Third-light $l_{3}$			& $0.057(13)$		& $0.107(39)$		& --					& --	 \\
\noalign{\smallskip}
\hline
\noalign{\smallskip}
Fractional radius, $r_1$ 		& $0.03838(37)$ 	& $0.0392(12)$		& $0.0380(5)$ 			&  $0.03829(29)$  \\ 
Fractional radius, $r_2$		& $0.06071(29)$ 	& $0.0606(8)$		& $ 0.0613(10)$		& $0.06076(28)$   \\ 
Eccentricity, $e$			& $0.1780(30)$ 	& $0.196(6)$		& $0.189(10)$ 			& $0.1821(51)$\\ 
Periastron Longitude, $\omega$ (\degr)	& $111.63(32)$ & $109.28(46)$		& $109.6(1.0)$		& $110.73(78)$ \\ 
\noalign{\smallskip}
\hline
\end{tabular}
\tablefoot{\tablefoottext{*}{Calculated from the values given by {\citet{1988A&A...196..128A}} for $r_ A$ and $r_B$.}}
\end{table*}

\section{Absolute parameters}
\label{sec:massAndRadii}

As mentioned in Sec. \ref{sec:Intro}, \citet{2009MNRAS.400..969H} obtained high precision radial velocity measurements for AI Phe and combined these with ASAS photometric measurements to derive the masses and radii of the two components of AI Phe. The values they used for the semi-amplitude velocities $K_{1}$ and $K_{2}$ are $51.36(3)$  \mbox{km s$^{\rm -1}$} and $49.11(2)$ \mbox{km s$^{\rm -1}$} respectively.

\begin{table} 
\caption{Absolute parameters for AI Phe parameters calculated using lightcurve parameters from this work and spectroscopic values from \citet{2009MNRAS.400..969H}, with comparisons to previous work.}
\label{tab:AIPheMassRadiiData}
{\centering{
\begin{tabular}{@{}l r r r }
\hline\hline
\noalign{\smallskip}
Parameter				& \multicolumn{1}{c}{This work} & \multicolumn{1}{c}{He\l{}minak} & \multicolumn{1}{c}{Andersen} \\
					& 						& \multicolumn{1}{c}{et al. (2009)} & \multicolumn{1}{c}{et al. (1988)} \\
\noalign{\smallskip}
\hline
\noalign{\smallskip}
$P$ (days)					& $24.592483$		& $24.59241$		& $24.592325$ 	\\
Error in $P$					& $(17)$			& $(8)$			& $(8)$ 			\\
$K_{1}$ \mbox{(km s$^{\rm -1})$}	& \multicolumn{2}{c}{$51.16(3)$} 		& $50.90(8)$  	\\
$K_{2}$ \mbox{(km s$^{\rm -1})$}	& \multicolumn{2}{c}{$49.11(2)$} 		& $49.24(7)$    \\
$q$							& $1.0417(7)$		& $1.0418(8)$ 		& $1.034(2)$  	 \\
$M_{1}\sin^{3}i$ (M$_{\sun}$)		& $1.1961(37)$ 	& $1.1922(30)$\parbox{0pt}{$^*$}	& $1.194(4)$ \\
$M_{2}\sin^{3}i$ (M$_{\sun}$)		& $1.2460(39)$	 	& $1.2421(32)$\parbox{0pt}{$^*$} 	& $1.234(5)$   \\
$e$							& $0.1821(51)$		& $0.187(4)$		& $0.188(2)$	\\
$\omega $(\degr) 				& $110.73(78)$		& $110.1(9)$		& $109.9(6)$	\\
$i$ (\degr)						& $88.502(39)$		& $84.4(5)$		& $88.45(5)$	\\
\noalign{\smallskip}
\hline
\noalign{\smallskip}
$M_{1}$ (M$_{\odot}$)			& $1.1973(37)$		& $1.2095(44)$\parbox{0pt}{$^*$}		& $1.1954(41)$	\\ 
$M_{2}$ (M$_{\odot}$)			& $1.2473(39)$		& $1.2600(46)$\parbox{0pt}{$^*$}		& $1.2357(45)$	\\
$R_{1}$ (R$_{\odot}$)			& $1.835(14)$		& $1.82(5)$						& $1.816(24)$	\\
$R_{2}$ (R$_{\odot}$)			& $2.912(14)$		& $2.81(7)$						& $2.930(48)$  \\
\noalign{\smallskip}
\hline
\end{tabular}}}
\tablefoot{\tablefoottext{*}{These errors have been recalculated using \textsc{jktabsdim} as the quoted errors have been under-estimated.}}
\end{table}
The WASP photometry provides more complete lightcurves for AI Phe than the ASAS data, and therefore it has been possible to obtain the lightcurve parameters to a higher precision. Using the weighted means for $r_1$, $r_2$, $e$ and $i$ from the Table~\ref{tab:AIPheParamData} and the semi-amplitude velocities from \citet{2009MNRAS.400..969H}, masses and radii of the stars within AI Phe have been calculated and are shown in Table~\ref{tab:AIPheMassRadiiData}. \textsc{jktabsdim}\footnote{http://www.astro.keele.ac.uk/jkt/codes/jktabsdim.html} was used for this, as were the constants as suggested by \citet{2010A&ARv..18...67T}. Table~\ref{tab:AIPheMassRadiiData} also contains results from \citet{1988A&A...196..128A} and \citet{2009MNRAS.400..969H} for comparison. Note that the uncertainties of $M_{1,2}\sin^{3}i$ and the masses quoted by \citeauthor{2009MNRAS.400..969H} have been recalculated using \textsc{jktabsdim}. Their uncertainties have been underestimated somewhat as their quoted uncertainties could not be reproduced with \textsc{jktabsdim}, despite using all values quoted in their paper. With the uncertainties in $K_{1}$ and $K_{2}$ being so small, the uncertainty in the eccentricity has become the largest source of error in the masses. Overall, the values obtained in this work for $M_{1,2}\sin^{3}i$ do have larger uncertainties than those of \citeauthor{2009MNRAS.400..969H}, which is due to the greater uncertainty in the eccentricity, but with the improved accuracy of the inclination, the uncertainties in the masses has been reduced to 0.31\% for both $M_{1}$ and $M_{2}$. While $M_{1}$ is consistent with the value found by \citet{1988A&A...196..128A}, there is almost a 3-$\sigma$ difference in $M_{2}$. This stems from different $M_{2}\sin^{3}i$, and so it is also seen in the values from \citeauthor{2009MNRAS.400..969H}. 

There has also been a significant reduction in the uncertainties associated with the radii. $R_1$ is known to a precision of 0.76\% while $R_2$ is known to a precision of 0.48\%, a reduction from 1.3\% and 1.6\% respectively \citep{1988A&A...196..128A}. Although not a concern for the radii presented here, below $0.1\%$ precision, uncertainties in the constant used for the solar radius, and the definition of a star's radius needs to be considered \citep{2010A&ARv..18...67T}.

The biggest contribution to the uncertainties of the masses is now the eccentricity. Further reduction in the uncertainties of the masses would require the orbit of AI Phe to be better understood, and an understanding of the variability shown in $e$ and $\omega$.

\section{Implication for models}
\label{sec:models}

The increase in precision associated with the masses and radii of AI Phe will provide tighter constraints on the models that can provide an estimate of the age of the system.  As such, a MCMC method has been used to estimate the age of AI Phe for a number of grids of models, with varying mixing length, $\alpha_{ml}$ and helium abundance, {$Y$}. The grids of models have been produced with the \textsc{garstec} stellar evolution code \citep{2008Ap&SS.316...99W} and methods used to calculate the grids are described by \cite{2015A&A...575A..36M} and \cite{2013MNRAS.429.3645S}. The MCMC method is very similar to the method described by \cite{2015A&A...575A..36M}, however there are a number of differences, which are described in more detail in Sects.~\ref{subsec:InData} and \ref{subsec:ageEstimates}. Most notably, the code will attempt to fit the two stars of AI Phe to the same isochrone instead of fitting each star individually. The priors have also been modified slightly in order to accommodate the new observed values for both stars.

{\textsc{garstec}} uses the \cite{1990sse..book.....K} mixing length theory for convection, where $\alpha_{ml} = 1.78$ produces the observed properties of the Sun assuming the composition given by \citet{1998SSRv...85..161G}, an initial solar helium abundance \mbox{$Y_{\sun}=0.26626$}, and initial solar metallicity \mbox{$Z_{\sun}=0.01826$}. The initial solar composition corresponds to an initial iron abundance of [Fe/H]$_{\rm i} = +0.06$, due the effects of microscopic diffusion.  OPAL radiative opacities of \cite{1996ApJ...464..943I} are used and are complemented by molecular opacities from \cite{2005ApJ...623..585F} at low temperatures.

Convective mixing is treated as a diffusive process, where the diffusion coefficient at each point is given by $D_{\rm c}= \frac{1}{3} \alpha_{\rm ml} H_{\rm P} v_{\rm c}$. $\alpha_{\rm ml} H_{\rm P}$ is the local mixing length, and $v_{\rm c}$ is the convective velocity determined from mixing length theory. Overshooting is included by extending the mixing region beyond the formal Schwarzschild boundary with an exponentially decaying diffusion coefficient, given as $D = D_{0} \exp{(- 2z / (f h_{\rm p}))}$. $D_{0}$ is the diffusion coefficient inside the convective border, $f = 0.020$ is a free parameter defining the scale of overshooting and
\begin{equation}
h_{\rm p} = H_{\rm P} \times {\rm min} \left[1, \left( \frac{\Delta R_{\rm CZ}}{H_{\rm P}} \right)^{2}  \right].
\end{equation}
$\Delta R_{\rm CZ}$ is the thickness of the convective core, and $H_{\rm P}$ is the pressure scale height. The definition of $h_{\rm p}$ ensures that for small convective regions, (particularly small convective cores where $\Delta R_{\rm CZ} < H_{\rm P}$) the overshooting region is geometrically limited to a fraction of the size of the convective region \citep{2010ApJ...718.1378M}. If $R_{\rm CZ} > H_{\rm P}$ the geometric limit does not play a role and the overshooting region amounts to $\sim 0.25 H_{\rm P}$, for convective cores in the main sequence. For stars in the range 1.2 - 1.3~${\rm M_\odot}$, when a convective core develops towards the end of the main sequence, the geometric cut effectively limits the size of the overshooting region to $\sim 0.05 H_{\rm P}$.

Atomic diffusion of all atomic species is included by solving the multi-component flow equations of \cite{1969fecg.book.....B}, according to the method of \cite{1994ApJ...421..828T}. Extra macroscopic mixing below the convective envelope is also included. It follows the parametrisation given in \cite{2012ApJ...755...15V} and depends on the extension of the convective envelope. Radiative accelerations and stellar winds are not included in model calculations for main-sequence stars. These effects become more relevant as stellar mass is increased and limit the efficiency of atomic diffusion which, if acting alone, would yield metal abundances in the stellar surface much lower than observed.  For this reason, the efficiency of atomic diffusion is smoothly decreased in models from 1.25-1.35\,M$_{\sun}$ and completely switched off for higher masses. 

The model grids used cover six different mixing lengths, (1.22, 1.36, 1.50, 1.78, 2.04 and 2.32) with the helium abundance $Y_{\rm AI}$ fixed. An additional ten model grids have helium abundances that change by $\Delta Y$ from an initial value in increments of 0.01, whilst the mixing length is fixed. $\Delta Y$ covers a range from $-0.05$ to $0.05$. The initial value, $Y_{0} = 0.261\pm 0.007$ was calculated using
\begin{equation}
Y_{0}  = Y_{\rm BBN} + Z_{\rm AI}\frac{d Y}{d Z}
\label{eq:HelAbunEq}
\end{equation}
where \mbox{$Y_{\rm BBN} = 0.2485$} is the primordial helium abundance at the time of the big-bang nucleosynthesis \citep{2010JCAP...04..029S}, and \mbox{$ Z_{\rm AI} = 0.012\pm 0.007$} is the initial metal content of AI Phe. Diffusion has been considered with the chosen $Z_{\rm AI}$, however, the effect is almost negligible, making $ Z_{\rm AI}$ nearly identical to the surface $Z$ \citep{1988A&A...196..128A}. \mbox{$d Y/d Z$} is an assumed helium-to-metal enrichment calculated as \mbox{$d Y/d Z = (Y_{\sun} - Y_{\rm BBN})/Z_{\sun}$ = 0.984} using \mbox{$Z_{\sun}=0.01826$} and \mbox{$Y_{\sun}=0.26626$} \citep{2015A&A...575A..36M}. The value of \mbox{$d Y/d Z$} is very uncertain. Here, we have used a solar calibrated value, however \mbox{$d Y/d Z$} can change depending on where in the Galaxy you look and values in the literature can range from 0.5 to 5 \citep{2014EAS....65...99L,2010A&A...518A..13G}. \cite{2014EAS....65...99L} showed that increasing \mbox{$d Y/d Z$} from 2 to 5, decreases the turn off age of the star. The mass range $0.6\,M_{\sun}$ to $2.0\,M_{\sun}$ is covered by the model grids, in steps of $0.02\,M_{\sun}$, while the initial metallicity, [Fe/H]$_{\rm i}$ covers $-0.75$ to $-0.05$ in steps of $0.1$ dex and $-0.05$ to $+0.55$ in steps of $0.05$ dex.

\subsection{Input data}
\label{subsec:InData}

For the MCMC analysis, a vector of parameters \mbox{\vec{d}=($T_{1}$, $\rho_{1}$, $\rho_{2}$, $T_{\rm ratio}$, $M_{\rm sum}$, $q$, ${\rm [Fe/H]_{s}}$)} can be used to define the observed quantities for AI Phe. These parameters have been chosen because each quantity is determined by an independent feature of the data used in the analysis, with little or no dependence on the other parameters. $T_{\rm ratio}$ is the ratio of the effective temperatures given by $T_{2}/T_{1}$, $T_{1}$ and $T_{2}$ are the effective temperatures of the two stars, $\rho_{1}$ and $\rho_{2}$ are the average stellar densities of the two stars, $M_{\rm sum}$ is the sum of their masses, $q$ is the mass ratio given by $M_{2}/M_{1}$ and [Fe/H]$_{\rm s}$ is the observed surface metal abundance.

The mass ratio and sum of the masses were chosen over directly using the individual masses $M_{1}$ and $M_{2}$, as the individual masses have stronger correlations. The densities of the two stars (\mbox{$\rho_{1}=0.1935\pm0.0044\,\rho_{\sun}$} and \mbox{$\rho_{2}=0.0505\pm0.0007\,\rho_{\sun}$}) were calculated using 
\begin{equation}
\rho_{\rm n} = \frac{3 \pi}{GP^2(1+Q_{\rm n})}\left(\frac{a}{R_{\rm n}}\right)^{3}
\label{eq:stellarDensity}
\end{equation}
where $R$ is the radius for star $n=1,2$, $a$ is the semi-major axis of the orbit, $P$ is the orbital period, $G$ is Newton's gravitational constant \citep{2015A&A...575A..36M}. $Q_{\rm n}$ is a function of the mass ratio, where $Q_{1} = q$ and $Q_{2} = 1/q$. Equation \ref{eq:stellarDensity} allows the density to be calculated directly from values of $r_{1}$ and $r_{2}$ derived from the lightcurve analysis using Kepler's law, and independently from the mass estimates from the spectroscopic orbit. $T_{1}$ was taken to be \mbox{$6310\pm150$ \,K} \citep{1985ApJ...291..270V}.

The ratio of the stars' effective temperatures can be determined directly from the surface brightness ratio derived from the lightcurve analysis. The surface brightness ratio is related to the ratio of the eclipse depths in a totally-eclipsing binary, with very little dependence on the other parameters of the lightcurve. An approximate value of $T_{\rm eff}$ is needed for one of the stars in the binary, but this value has only a small effect on the derived $T_{\rm ratio}$. Our method uses \cite{1993yCat.6039....0K} model atmospheres and the profiles for numerous passbands, \citep{1990PASP..102.1181B, 1970AJ.....75..978C, 2010AJ....139.1628D} and is similar to the method described by \cite{2015A&A...578A..25M}. For each of the bands, Johnson $BVRI$, Str\"omgen $y$ and SDSS $r'$, a relationship was established between effective temperature and surface brightness for \mbox{$\log g=4.0$} and \mbox{$\log g =3.6$}. Interpolation between the values from the models for \mbox{$\log g=4.0$} and \mbox{$\log g=3.5$} was used in the case of $\log g=3.6$ as no model was available. Taking \mbox{$T_{1} =6310\pm150$ \,K} we used these relationships to find a value for $T_{2}$ that gave the measured average surface brightness ratio for each band. The average surface brightness ratios were calculated from the central surface brightness ratios for the {\em{BVRI}} and {\em{y}} passbands from \citet{1988A&A...196..128A} and the 85-mm central surface brightness ratio was used for SDSS $r'$ passband. \citet{1988A&A...196..128A} did not define their surface brightness ratios as average or central values. We have assumed the values are central surface brightness ratios because their lightcurve model uses the same methods as \textsc{ebop} and should therefore produce the same ratios as \textsc{ebop}. Following the example of \citet{1988A&A...196..128A}, limb darkening coefficients for the {\em{BVRI}} were taken from \citet{1984ApJ...282..748H} and limb darkening coefficients for {\em{y}} from \citet{1985A&AS...60..471W}. The values of $T_{\rm ratio}$ derived using this method for the different passbands were found to be consistent with each other. The weighted mean and standard error are $T_{\rm ratio} = 0.83\pm 0.01$, where the standard error estimate accounts for the uncertainties on both $J$ and $T_{1}$. There are surface brightness ratios available for the Johnson $U$ and Str\"omgen $uvb$ passbands in the \citet{1988A&A...196..128A} paper, however this method is less reliable in the bluer passbands as line-blanketing is more prevalent. These bands were therefore excluded from the determination of $T_{\rm ratio}$. Another consideration is the metallicity used by the models. The \cite{1993yCat.6039....0K} model atmospheres consider solar metallicity, instead of the measured value of AI Phe \citep[$-0.14\pm0.1$,][]{1988A&A...196..128A}. However, \citet{2015A&A...578A..25M} found that changing the metallicity by $\pm0.1$ dex changes the resulting $T_{2}$ by less than 10\,K, or $\approx 0.002$ in $T_{\rm ratio}$. As such, the resulting error is within the uncertainties of $T_{\rm ratio}$.

\subsection{Bayesian age estimates}
\label{subsec:ageEstimates}

The model parameters used to predict the observed data can be represented as \mbox{$\vec{m} = \left(\mbox{$\tau_{\rm sys}$}, \mbox{$M_{1}$}, \mbox{$M_{2}$}, \mathrm{[Fe/H]}_{\mathrm{i}}\right)$}, where $\mbox{$\tau_{\rm sys}$}$, $\mbox{$M_{1}$}$, $\mbox{$M_{2}$}$ and $\mathrm{[Fe/H]}_{\mathrm{i}}$ are the age, mass of star 1, mass of star 2, and initial metal abundance of the system, respectively. Due to diffusion and mixing processes occurring in the star during its evolution, the initial metal abundance, $\mathrm{[Fe/H]}_{\mathrm{i}}$ differs from the observed surface metal abundance, $\mathrm{[Fe/H]}_{\mathrm{s}}$.

The probability distribution function \mbox{$p(\vec{m}|\vec{d}) \propto {\cal L}(\vec{d}|\vec{m})p(\vec{m})$} was determined using a MCMC method. \mbox{${\cal L}(\vec{d}|\vec{m}) = \exp(-\chi^2/2)$} is used to estimate the likelihood of observing the data $\vec{d}$ for a given model $\vec{m}$, where
\begin{eqnarray} 
\chi^2   = & \left[\sum_{n=1,2}\frac{\left(\rho_{n} - \rho_{n,\mathrm{obs}}\right)^2}{\sigma_{\rho_{n}}^2} \right] 
+\frac{\left(T_{\mathrm{1}} -T_{\mathrm{1,obs}}\right)^2}{\sigma_{T_{1}}^2}
+ \frac{\left(T_{\rm ratio} - T_{\rm ratio, \mathrm{obs}}\right)^2}{\sigma_{T_{\rm ratio}}^2} \\
\noalign{\smallskip}
& + \frac{\left(M_{\rm sum} - M_{\rm sum, \mathrm{obs}}\right)^2}{\sigma_{M_{\rm sum}}^2}
+ \frac{\left(q - q_{\mathrm{obs}}\right)^2}{\sigma_{q}^2}  + \frac{\left(\mathrm{[Fe/H]}_{\mathrm{s}} - \mathrm{[Fe/H]}_{\mathrm{s,obs}}\right)^2}{\sigma_{\mathrm{[Fe/H]_{\mathrm{s}}}}^2} \nonumber.
\end{eqnarray}

\noindent Observed quantities are denoted with `obs' subscript and their standard errors are given by the appropriately marked $\sigma$. The probability distribution function \mbox{$p(\vec{m}) = p(\mbox{$\tau_{\rm sys}$})p(\mbox{$M_{1}$})p(\mbox{$M_{2}$})p(\mathrm{[Fe/H]}_{\mathrm{i}})$} is the product of the individual priors on each of the model parameters. The assumed prior on $\mathrm{[Fe/H]}_{\mathrm{i}}$ normally has little effect because this parameter is well constrained by the observed value of $\mathrm{[Fe/H]}_{\mathrm{s}}$ so a `flat' prior on $\mathrm{[Fe/H]}_{\mathrm{i}}$ is used, i.e., a uniform distribution over the model grid range. Although there is a prior on the age to keep it within the limits of 0\,--\,17.5\,Gyr, the age does not venture close to these limits as the other priors provide much tighter constraints on the age of the system. The code also offers the option to set a prior on the surface $\mathrm{[Fe/H]}$ for cases where there is no observed value of surface metallicity. As the $\mathrm{[Fe/H]}_{\mathrm{s}}$ is tightly constrained from observation, a flat prior of $-0.75<\mathrm{[Fe/H]}<0.55$ was used.

The Markov chain setup uses the same approach described by \cite{2015A&A...575A..36M}, in that a Markov chain of points $\vec{m}_i$ is created with the probability distribution \mbox{$p(\vec{m}|\vec{d})$} using a jump probability distribution $f(\Delta\vec{m})$. The generation of each trial point $\vec{m}^{\prime} = \vec{m}_i + \Delta\vec{m}$ is dictated by probability distribution, with the point being rejected if any of the model parameters are outside of the ranges set by the priors. If ${\cal L}(\vec{m}^{\prime}|\vec{d}) > {\cal L}(\vec{m}_i|\vec{d})$, the point is always included, and when ${\cal L}(\vec{m}^{\prime}|\vec{d}) < {\cal L}(\vec{m}_i|\vec{d})$ the point is included with probability ${\cal L}(\vec{m}^{\prime}|\vec{d})/{\cal L}(\vec{m}_i|\vec{d})$ (Metropolis-Hastings algorithm). $\vec{m}_{i+1} = \vec{m}^{\prime}$ if the trial point is accepted, or $\vec{m}_{i+1}=\vec{m}_{i}$ otherwise \citep{Metropolis1953,HASTINGS01041970}.

In comparison to \cite{2015A&A...575A..36M}, a different approach is used to find a starting point. Their approach takes the point with the lowest $\chi^2$ when a sample of points are randomly generated across the model grid. In our case, because the parameter-space is so well constrained, the starting point needed to be guided towards the parameter-space. The measured masses and metallicity are fixed and used to generate an evolutionary track for each star in the system. From there, each point along the tracks is searched to find the point with the lowest $\chi^2$, using 2\,000 steps in age. This is done separately for the main-sequence and post main-sequence stages, with the best-fit point from this process is used as an initial starting point.

For each parameter a step size is found such that $|\ln({\cal L}(\vec{m}_0|\vec{d}) - \ln({\cal L}(\vec{m}_0^{(j)}|\vec{d})| \approx 0.5$, where $\vec{m}_0^{(j)}$ denotes a set of model parameters that differs from $\vec{m}_0$ only in the value of one parameter, $j$. From there, a burn-in stage of 10\,000 steps is used to improve the initial set of parameters and to determine correlations between parameters by calculating a covariance matrix. The eigenvectors and eigenvalues of this matrix are used to determine a set of uncorrelated, transformed parameters,  $\vec{q} = (q_1, q_2, q_3, q_4)$, where each of the transformed parameters has unit variance \citep{2004PhRvD..69j3501T}. To estimate the probability distribution \mbox{$p(\vec{m}|\vec{d})$}, a second Markov chain of 1\,000\,000 steps is produced, using a Gaussian distribution for $f(\Delta\vec{q})$ with unit standard deviation for each of the transformed parameters and the most likely model parameters as a starting point. With the tight constraints given by the priors, some of the model grids have a severely restricted parameter-space and so required a large number of steps to ensure the space was explored. All chains were visually inspected and checked via a running mean to ensure suitable mixing.

\subsection{Model comparisons}
\label{subsec:ModelComparisons}

\begin{figure}
\resizebox{\hsize}{!}{\includegraphics{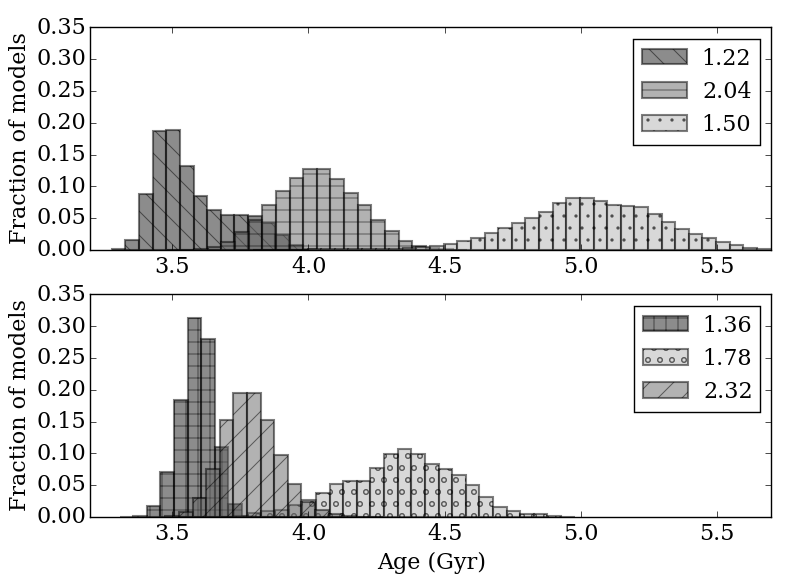}}
\caption{Age distributions of AI Phe obtained for six different values of mixing length, whilst helium abundance is held fixed at zero. Based on Markov chains of 1\,000\,000 steps.} 
\label{fig:mixLengthAges}
\end{figure}

\begin{figure}
\resizebox{\hsize}{!}{\includegraphics{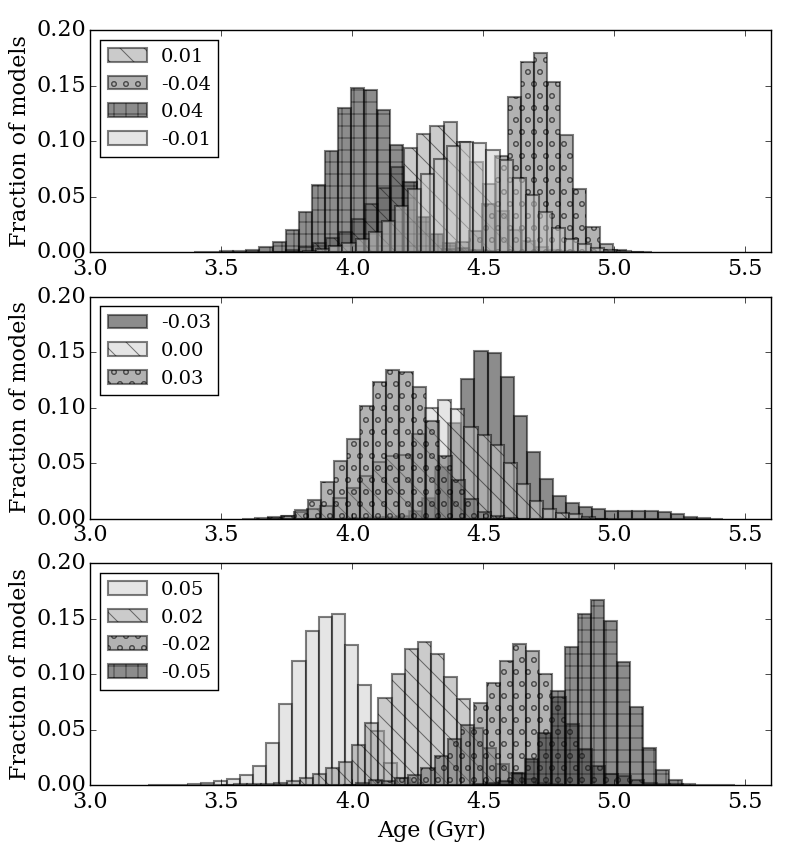}}
\caption{Age distributions of AI Phe obtained for different values of helium abundance whilst fixing the mixing length at 1.78. Based on Markov chains of 1\,000\,000 steps.} 
\label{fig:HeAbunAges}
\end{figure}

\begin{table*}
\caption{ {Age and parameters from the best fitting model from a 1\,000\,000-step bayesian age fitting method, for model grids with different mixing lengths and helium abundances. The mean and standard deviation of the resulting age distribution for each model grid is also shown.}}
\label{tab:AgeParams}
\centering
\begin{tabular}{c c c c c r r r r r r r r}
\hline\hline
\noalign{\smallskip}
${\alpha_{\rm ml}}$ & $\Delta Y$ & \multicolumn{1}{c}{$\tau_{\rm best}$} &  \multicolumn{1}{c}{$\tau_{\rm mean}$} & \multicolumn{1}{c}{$\sigma_{\tau_{\rm mean}}$} &  \multicolumn{1}{c}{$M_{1}$} &  \multicolumn{1}{c}{$M_{2}$} &  \multicolumn{1}{c}{${\rm [Fe/H]_{i}}$} &  \multicolumn{1}{c}{$T_{1}$} &  \multicolumn{1}{c}{$T_{2}$} &  \multicolumn{1}{c}{$\rho_{1}$} &  \multicolumn{1}{c}{$\rho_{2}$} 	&  \multicolumn{1}{c}{$\chi^{2}$}  \\

 & & \multicolumn{1}{c}{(Gyr)} &  \multicolumn{1}{c}{(Gyr)} & \multicolumn{1}{c}{(Gyr)} &  \multicolumn{1}{c}{$(M_{\sun})$} &  \multicolumn{1}{c}{$(M_{\sun})$} &  &  \multicolumn{1}{c}{(K)} &  \multicolumn{1}{c}{(K)}  &  \multicolumn{1}{c}{($\rho_{\sun}$)} & \multicolumn{1}{c}{($\rho_{\sun}$)} 	&  \\
\noalign{\smallskip}
\hline
\noalign{\smallskip}
1.22	& $0.00$	& $3.47$	& 3.58\parbox{0pt}{$^*$} & 0.14 & 1.1963 & 1.2467 & $-0.44$ & 6567 & 5685 & 0.1873 & 0.0499 & 32.8  \\
1.36	& $0.00$	& $3.60$	& 3.59 & 0.06 & 1.1948 & 1.2456 & $-0.39$ & 6484 & 5459 & 0.1807 & 0.0504 & 21.7	   \\
1.50	& $0.00$	& $5.03$	& 5.03 & 0.25 & 1.1956 & 1.2460 & $0.03$ & 5934 & 4804 & 0.1862 & 0.0507 & 16.4	  \\
1.78	& $0.00$	& $4.39$	& 4.34 & 0.20 & 1.1974 & 1.2472 & $-0.14$ & 6257 & 5100 & 0.1937 & 0.0506 & 2.4	  \\
2.04	& $0.00$ 	& $4.02$	& 4.03 & 0.15 & 1.1988 & 1.2481 & $-0.25$ & 6476 & 5223 & 0.2001 & 0.0504 & 6.5	  \\
2.32	& $0.00$ 	& $3.77$	& 3.79 & 0.10 & 1.2002 & 1.2491 & $-0.34$ & 6658 & 5514 & 0.2053 & 0.0503 & 20.0	  \\

\noalign{\smallskip}
\hline
\noalign{\smallskip}
1.78	& $-0.05$	& $4.95$	& 4.92 & 0.12 & 1.2025 & 1.2506 & $-0.40$ & 6373 & 5208 & 0.2137 & 0.0502 & 41.0	 \\
1.78	& $-0.04$	& $4.71$	& 4.70 & 0.11 & 1.2010 & 1.2494 & $-0.39$ & 6394 & 5219 & 0.2094 & 0.0503 & 28.5	 \\
1.78	& $-0.03$	& $4.52$	& 4.56 & 0.17 & 1.1997 & 1.2484 & $-0.36$ & 6397 & 5208 & 0.2062 & 0.0504 & 20.0	 \\
1.78	& $-0.02$	& $4.63$	& 4.63 & 0.18 & 1.1997 & 1.2489 & $-0.23$ & 6292 & 5129 & 0.2023 & 0.0505 & 9.1 \\
1.78	& $-0.01$	& $4.47$	& 4.45 & 0.20 & 1.1983 & 1.2478 & $-0.20$ & 6288 & 5124 & 0.1977 & 0.0505 & 4.1	  \\
1.78	& $0.00$	& $4.39$	& 4.34 & 0.20 & 1.1974 & 1.2472 & $-0.14$ & 6257 & 5100 & 0.1937 & 0.0506 & 2.4	 \\
1.78	& $0.01$	& $4.34$	& 4.31 & 0.17 & 1.1970 & 1.2469 & $-0.07$ & 6221 & 5071 & 0.1918 & 0.0506 & 3.2 \\
1.78	& $0.02$ 	& $4.27$ 	& 4.26 & 0.17 & 1.1967 & 1.2468 & $-0.01$ & 6194 & 5048 & 0.1905 & 0.0506 & 5.2 \\
1.78	& $0.03$	& $4.17$	& 4.16 & 0.14 & 1.1964 & 1.2466 & $0.05$ & 6178 & 5032 & 0.1892 & 0.0506 & 8.2	 \\
1.78	& $0.04$	& $4.06$	& 4.04 & 0.14 & 1.1962 & 1.2465 & $0.10$ & 6165 & 5019 & 0.1884 & 0.0507 & 11.7	 \\
1.78	& $0.05$	& $3.91$	& 3.90 & 0.13 & 1.1963 & 1.2467 & $0.15$ & 6165 & 5015 & 0.1876 & 0.0507 & 15.5	\\
\noalign{\smallskip} 
\hline
\end{tabular}
\tablefoot{\tablefoottext{*}{The is a noticeable difference between the mean and best-fit ages for this model grid, as this model grid produced a bimodal distribution and as such the age distribution does not match a gaussian profile (see Fig. \ref{fig:mixLengthAges}).}}
\end{table*}

Using the Bayesian method described in Sects.~\ref{subsec:InData} and \ref{subsec:ageEstimates} an age distribution has been produced for a number of model grids, with different mixing lengths $\alpha_{\rm ml}$ and helium abundance $Y$. In total, distributions were produced for six mixing lengths  (1.22, 1.36, 1.50, 1.78, 2.04 and 2.32) with $\Delta Y$ held fixed at zero (see Fig.~\ref{fig:mixLengthAges}) and an additional ten distributions were produced for helium abundances, where $\Delta Y$ ranged from $-0.05$ to $0.05$ in increments of 0.01, with the mixing length held fixed at 1.78 (Fig. \ref{fig:HeAbunAges}). Table~\ref{tab:AgeParams} contains the parameters from the model with the lowest $\chi^2$ for each model grid, alongside the mean $\tau_{\rm mean}$ and standard deviation $\sigma_{\tau_{\rm mean}}$ of the age distributions produced by the 1\,000\,000-step chain. In some cases the age distributions produced bimodal distributions which do not fit a Gaussian profile, meaning there is a noticeable difference between the best-fit age $\tau_{\rm best}$ and $\tau_{\rm mean}$. The first section of the table contains the model grids where ${\alpha_{\rm ml}}$ is varied, and the second section contains model grids where the helium abundance, $Y_{\rm AI}$, is varied by $\Delta Y$ from the initial value $Y_{0}$. 

The $\chi^2$ values in Table~\ref{tab:AgeParams} show that some of the model grids fit in the parameter-space much better than others. In terms of the mixing lengths, values of 2.04 and 1.78 are favoured, with 1.78 producing the lowest $\chi^2$. Using the 16th, 50th and 84th percentiles, a mixing length of 1.78 gives an age of $4.35^{+0.23}_{-0.19}$ Gyr. As the mixing length in the model grid is increased, the best-fit model tends to increase the mass of the two stars.  Comparing evolutionary tracks for star 2 (the sub-giant) at a fixed age, mass and metallicity but with increasing mixing length, results in larger densities, as the stars are more compact \citep{2014EAS....65...99L}. However, with the tight observational constraints on $\rho_{2}$, the models attempt to improve the overall $\chi^{2}$ by increasing $M_{2}$ and by decreasing the metallicity. Increasing $M_{2}$ results in a star with a larger radius with the same density, but also reduces the age of the sub-giant and therefore the system. The effect on the age can be seen in the mean and best-fit ages as $\alpha_{\rm ml}$ is changed from 1.50 to 2.32. Decreasing the metallicity allows the system to evolve to the same evolutionary stage faster.

The $\alpha_{\rm ml}=1.22$ model grid does not follow the mass trend mentioned above, because the mass distribution is bi-modal, corresponding to two different evolutionary stages. The most dominant part of the distribution places the secondary very early in the contraction phase before the ascent to the red-giant branch, with a larger effective temperature. The less prominent part of the distribution, places the secondary further into the contraction phase. A similar explanation can be used to explain the sudden change in best-fit age between $\alpha_{\rm ml}=1.36$ and $\alpha_{\rm ml}=1.50$. For the $\alpha_{\rm ml}=1.36$ model, the secondary component sits firmly in the contraction phase, whereas for $\alpha_{\rm ml}\geqslant 1.50$, the secondary sits at the base of the red-giant branch. 

Increasing the helium abundance used in the model grid has decreased the best-fit age of the system, with the models preferring to use smaller masses and cooler effective temperatures for both stars. The best-fit values for $\rho_{2}$ show very little variation with helium abundance, changing by less than 1\%. Meanwhile, $\rho_{1}$ shows a much larger variation of 14\%. The small uncertainty in $M_{2}$ tightly constraints the age of the sub-giant, allowing little variation in $\rho_{2}$. This is not the case with the main-sequence star. In order to find a model for the main-sequence star that fits the age determined by the sub-giant, parameters such as its effective temperature and density need to be varied significantly from their observed value for some values of the helium abundance. This is reflected in their much larger $\chi^2$ values.

Looking at the $\chi^2$ for the model grid with varying $\Delta Y$ in Table~\ref{tab:AgeParams}, the preferred models have a helium abundance that is closer to the initial value, effectively excluding values where $\Delta Y\leqslant-0.03$ and $\Delta Y\geqslant0.04$. The preferred model is where $\Delta Y=0.0$, meaning $Y_{\rm AI}=0.261$. Using $\Delta \chi^2 = 1$ to define a 68.3\% confidence interval on $\Delta Y$, based on the projection into the $\alpha_{\rm ml}$-$\Delta Y$ parameter space \citep{1992nrfa.book.....P}, the models for $\Delta Y=-0.01$ and $\Delta Y =0.02$ fall just outside this interval. A $\Delta \chi^2 = 2.71$ defines a 90\% confidence interval and covers the models $-0.02<\Delta Y <0.03$. We find \mbox{$Y_{\rm AI} = 0.26^{+0.02}_{-0.01}$} for a fixed mixing length of $1.78$, giving an age of \mbox{$4.39\pm0.32$\,Gyr} using $\tau_{\rm best}$ from the best-fit model. The uncertainty is estimated by directly adding $\sigma_{\tau_{\rm mean}}$ from $\Delta Y=0$ ($0.20\,$Gyr) and a systematic uncertainty of 0.12\,Gyr (from not knowing $Y_{\rm AI}$).
For comparison, \cite{2010A&ARv..18...67T} found an age of 4.1\,Gyr using experimental Victoria models \citep{2006ApJS..162..375V} and 5.0\,Gyr from Yonsei-Yale models \citep{2004ApJS..155..667D}, although no uncertainties are given. \cite{0004-637X-776-2-87} found the age of two components of AI Phe separately using an updated version of the Yale Rotational stellar Evolution Code (YREC). They find an age of $4.44\pm0.08\,$Gyr for the hotter component, and an age of $4.54\pm0.02\,$Gyr for the cooler component meaning our value is consistent with their ages.

Comparisons between mixing lengths are quite difficult because of the different approaches to used to calibrate the parameter, and how the mixing length is included in the models. \citet{1988A&A...196..128A} used a fixed mixing length of 1.50, which was calibrated by the producing a $1.0\,M_{\sun}$ star with $1.0\,R_{\sun}$ and the solar age of $4.7\,{\rm Gyr}$. They used interpolation between a solar model and a model with a $Z$ of 0.01 to find the age and helium abundance of AI Phe. \cite{0004-637X-776-2-87} used a similar solar calibration method but found the mixing-length was $\alpha = 1.875$, and \cite{2010A&ARv..18...67T} noted that at the time the Victoria models \citep{2006ApJS..162..375V} that had not been fully calibrated. Therefore, we note that our best-fit mixing length is the same as the solar value for our set of models. This is consistent with what was found by \citet{1988A&A...196..128A}. 

\begin{figure}
\resizebox{\hsize}{!}{\includegraphics{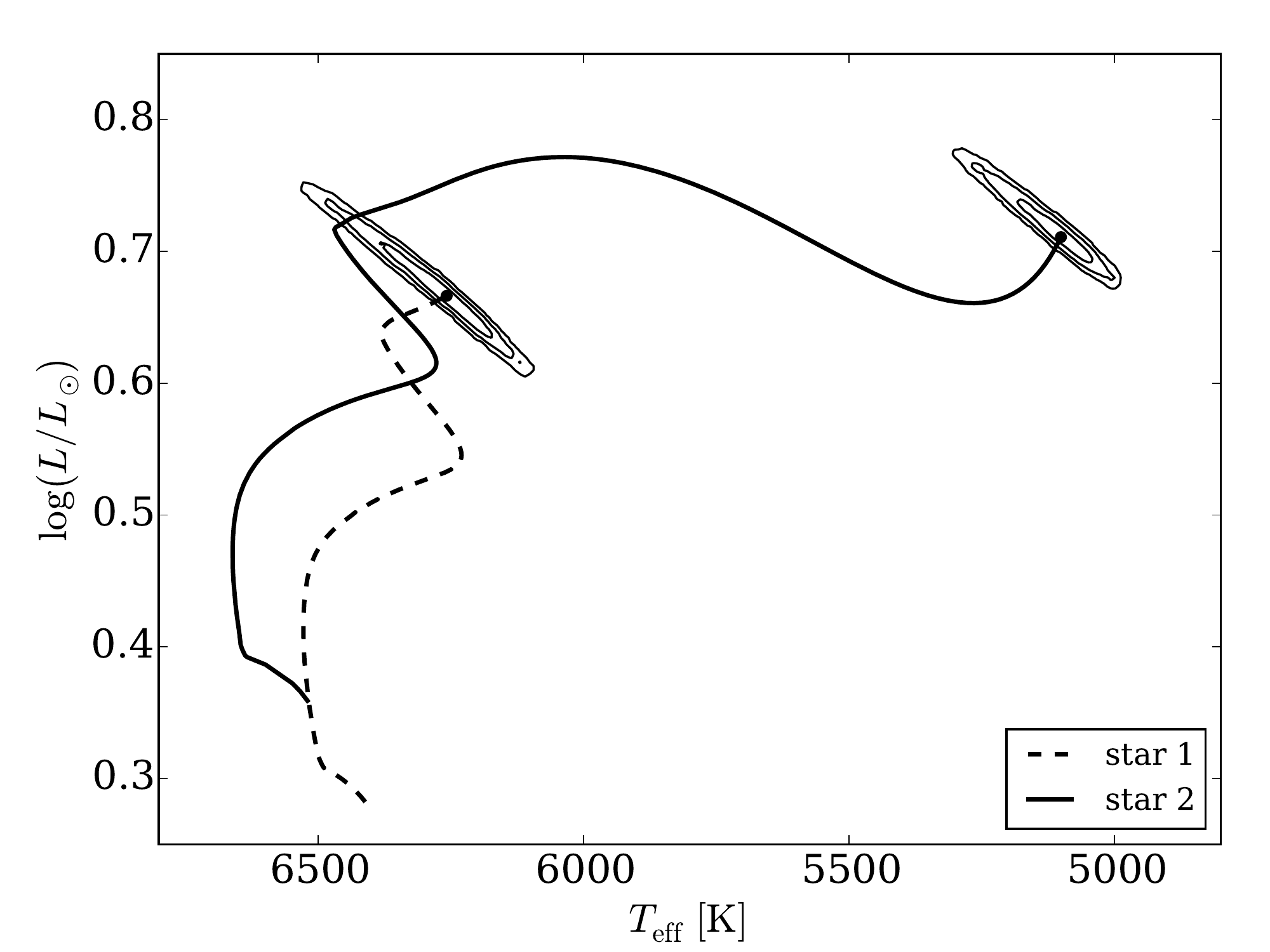}}
\caption{Theoretical tracks for the two components of AI Phe consistent with the best-fit age for the model grid with $\alpha_{\rm ml}=1.78$ and $\Delta Y=0.0$. 1$\sigma$, 2$\sigma$ and 3$\sigma$ confidence contours from the Bayesian analysis are shown.} 
\label{fig:LogLumT}
\end{figure}

\begin{figure}
\resizebox{\hsize}{!}{\includegraphics{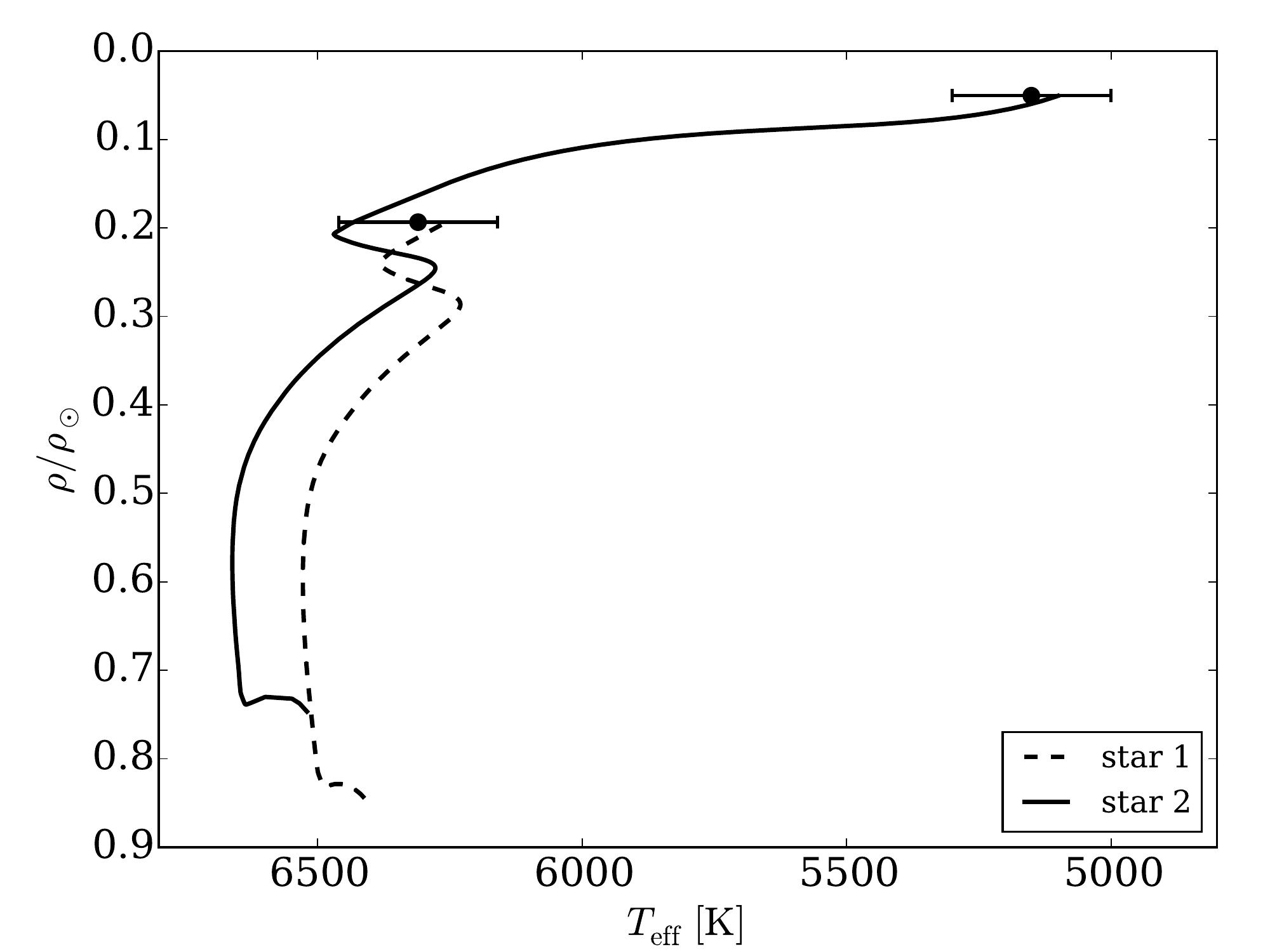}}
\caption{Observed density and effective temperature \citep{1992ApJS...79..123M} for the two components of AI Phe, for the theoretical tracks consistent with the best-fit age for the model grid with $\alpha_{\rm ml}=1.78$ and $\Delta Y=0.0$. The uncertainty in the densities are smaller than the markers.} 
\label{fig:densT}
\end{figure}

Figures \ref{fig:LogLumT} and \ref{fig:densT} show the theoretical tracks for the two components of AI Phe which result in the best-fit age for the model grid with $\alpha_{\rm ml}=1.78$ and $\Delta Y=0.0$. In Fig. \ref{fig:densT}, we have plotted the observed densities calculated in this work and the effective temperatures from \citet{1992ApJS...79..123M}.

\section{Discussion}
\label{sec:Discuss}

While the ephemeris in Equation~\ref{eq:ephemeris} is suitable for the WASP photometry presented here, it is not suitable for describing the orbit on a long-term basis. It has been shown that the period of AI Phe does not follow a linear ephemeris, with currently available times of primary minimum suggesting it may be more quadratic in nature. The exact cause is not currently known and further work will be needed to determine its nature. Insufficient coverage of the secondary eclipse has prevented us testing how the timings of the secondary eclipse has been affected by these period changes. If the deviations in timings are common to both the primary and secondary minima, it may suggest a third body is involved, or if the deviations are in opposite directions then it may suggest apsidal motion is the cause. Looking for possible trends in the variation in the eccentricity and longitude of periastron with currently available measurements, proved inconclusive. As the eccentricity is now the largest uncertainty in the determination of the masses, understanding the orbit of AI Phe will be essential if the precision of the masses are to be improved further.
 
The WASP detrending functions have little impact on the determined fractional radii and therefore the radii. Neither did the slight inconsistencies in the determined value of $e$ and $\omega$. In the first case, parameters determined from the 200-mm data have a greater sensitivity to the detrending functions in comparison to the 85-mm data. It is thought to be related to slight variations in the transmission profile of the two filters used, resulting in  small differences in the surface brightness and third-light of the 200-mm data. In the second case, the well-defined contact points of the primary eclipse, and the small, fourth-root dependence on $k$ allow the robust measurements of $r_{1}$ and $r_{2}$ despite the issues measuring the secondary eclipse.

In the models used for estimating the age of AI Phe, we have assumed that the mixing length and the helium abundance is the same for both of the stars. As the stars are binary components of the same age, they will have formed together from similar material and so using the same initial helium abundance for both stars is valid. As for the mixing length, investigations by \cite{2013ApJ...769...18T} found that as stars ascend the giant branch, $\alpha_{\rm ml}$ remains constant along the track. However, their calibrations of $\alpha_{\rm ml}$ for different $T_{\rm eff}$ and $\log g$ using radiation hydrodynamics simulations suggest that the cooler component of AI Phe should have a slightly larger mixing length, due to a difference of 0.4 in $\log g$ between the two components and a temperature difference of $\approx 1200\,{\rm K}$. This would mean a slightly smaller radius and larger $T_{\rm eff}$ for the star \citep{0004-637X-776-2-87}. The age of the system is largely constrained by the prior on the mass of $M_{2}$, so a small change in the radius and effective temperature would have little impact on the estimated age.

The models presented here have not explored any potential correlations between the mixing length and helium abundance, due to nature of the model grids. This means a better solution, other than the best-fit solutions in Table~\ref{tab:AgeParams}, could exist elsewhere in the $\alpha_{\rm ml}$-$\Delta Y$ plane if both $\alpha_{\rm ml}$ and $\Delta Y$ where varied simultaneously. Once we have obtained accurate parameters for four other binary systems, identified using WASP photometry, exploring the $\alpha_{\rm ml}$-$\Delta Y$ plane is something we intend to do.

Another consideration not explored here is the impact of the convective overshooting on the determined mixing length parameter and helium abundance. The components of AI Phe would provide an important test for convective overshooting as they sit on the mass boundary where stars start to develop convective cores \citep{2014EAS....65...99L}. For a star near the end of the main-sequence star, \cite{2016A&A...587A..16V} found that an uncertainty of $\pm1$ in the helium-to-metal enrichment ratio, could change the overshooting parameter by $\pm 0.03$, while a variation in $\alpha_{\rm ml}=0.24$ could change the overshooting parameter from $-0.03$ to $+0.07$. However, they used a different implementation of the overshooting region to the description used in this paper, making it difficult to translate the effects to AI Phe. The effects of convective overshooting is something that should be explored, and tighter constraints on the helium abundance and mixing length should help pin down this parameter.

\section{Conclusion}
\label{sec:Conc}

Using WASP photometry, the radii of components in AI Phe were found to be \mbox{$R_{1} = 1.835\pm0.014\,{\rm R}_{\sun}$} and \mbox{$R_{2} = 2.912\pm0.014\,{\rm R}_{\sun}$} with the uncertainties in the measurements reduced to 0.76\% for $R_{1}$ and 0.48\% for $R_2$. The masses were found to be \mbox{$M_{1} = 1.1973\pm0.0037\,{\rm M}_{\sun}$} and \mbox{$M_{2} = 1,.2473\pm0.0039\,{\rm M}_{\sun}$}, with their uncertainties reduced to 0.31\% for both components. The eccentricity is now the largest source of uncertainty in the masses.

Reducing the uncertainties on the masses and radii of the two stars in AI Phe, has meant that it has been possible to obtain the age of system, \mbox{$4.39\pm0.32$\,Gyr}, with greater precision than had been achieved by \citet{1988A&A...196..128A}. From the different model grids that were tested, a mixing-length of 1.78 and a initial helium abundance of $Y_{\rm AI}= 0.26^{+0.02}_{-0.01}$ gave the best-fit with the priors set by observations. 

Overall, with the new masses and radii obtained with the WASP photometry and spectroscopic orbit data of \citet{2009MNRAS.400..969H}, it has been shown that it is possible to constrain the mixing length and helium abundance of AI Phe. It is still possible to improve the results further by improving the eccentricity measurement and by exploring models with simultaneous variations in mixing length and helium abundance. However, additional data and new models would be required for this. In order to compare these results in terms of mixing length and helium abundance to other binary systems, it will be necessary to obtain absolute parameters for other binary systems to a similar level of precision as obtained here. 

\begin{acknowledgements}
WASP-South is hosted by the South African Astronomical Observatory and we are grateful for their ongoing support and assistance. Funding for WASP comes from consortium universities and from the UK's Science and Technology Facilities Council. J.A.K.-K. and D.F.E are also funded by the UK's Science and Technology Facilities Council. A.M.S. acknowledges support from grants SGR1458 (Generalitat de Catalunya), ESP2014-56003-R and ESP2015-66134-R (MINECO). This research has made use of "Aladin sky atlas" developed at CDS, Strasbourg Observatory, France and NASA's Astrophysics Data System.
\end{acknowledgements}

\bibliographystyle{aa} 
\bibliography{AIPhe_paper}
\end{document}